\documentclass{myarticle}

\newsavebox{\fminipagebox}
\NewDocumentEnvironment{fminipage}{m O{\fboxsep}}
{\par\kern#2\noindent\begin{lrbox}{\fminipagebox}
		\begin{minipage}{#1}\ignorespaces}
		{\end{minipage}\end{lrbox}%
	\makebox[#1]{%
		\kern\dimexpr-\fboxsep-\fboxrule\relax
		\fbox{\usebox{\fminipagebox}}%
		\kern\dimexpr-\fboxsep-\fboxrule\relax
	}\par\kern#2
}

\newcommand{\R}{\mathbb{R}}

\newcommand{\N}{\mathbb{N}}

\newcommand{\half}{\frac{1}{2}}

\newcommand{\squared}{^{2}}

\newcommand{\intzerot}{\int_{0}^{t}}

\newcommand{\timeHorizon}{T}
\newcommand{\timeWindow}{{[0,\timeHorizon]}}

\newcommand{\derivative}{^{\prime}}

\newcommand{\Lone}{L^{1}}

\newcommand{\Prob}{{{P}}}

\newcommand{\Expectation}{{{E}}}

\newcommand{\sigmaAlgebra}{\mathfrak{F}}

\newcommand{\stochasticBase}{\big(\Omega,\sigmaAlgebra = (\sigmaAlgebra_t)_t , \Prob  \big)}
\newcommand{\filtrationF}{\mathfrak{F}}
\newcommand{\internalHistory}{\filtrationF}

\newcommand{\compensator}{\Lambda}

\def\one{\mbox{1\hspace{-4.25pt}\fontsize{12}{14.4}\selectfont\textrm{1}}}

\newcommand{\numEventTypes}{d_{E}}
\newcommand{\numStates}{d_{S}}
\newcommand{\arrivalTimes}[1][e]{T^{#1}}
\newcommand{\event}[1][n]{E_{#1}}
\newcommand{\countingProc}[1][e]{N_{#1}}
\newcommand{\multiCountingProc}{N}
\newcommand{\sdHawkesPair}{(\multiCountingProc,\stateVariable)}
\newcommand{\groundProc}{N_{\mathfrak{g}}}
\newcommand{\hybridHawkes}{\tilde{N}}
\newcommand{\hybridIntensity}{\tilde{\intensity[]}}
\newcommand{\intensity}[1][e]{\lambda_{#1}}
\newcommand{\baseRate}{\nu}
\newcommand{\hawkesKernel}{\kappa}

\newcommand{\impCoef}{\alpha}
\newcommand{\decCoef}{\beta}
\newcommand{\dirichletParam}{\gamma}
\newcommand{\dirparam}{\dirichletParam}
\newcommand{\clusteringRate}{a}
\newcommand{\ordersizing}{c}
\newcommand{\eone}{e\derivative}
\newcommand{\subscriptee}{_{\eone,\, e}}
\newcommand{\xone}{x\derivative}
\newcommand{\subscriptexe}{_{\eone,\, \xone,\, e}}

\newcommand{\stateSpace}{\mathcal{S}}
\newcommand{\deflationarySpace}{\stateSpace^{-}}

\newcommand{\inflationarySpace}{\stateSpace^{+}}

\newcommand{\intensityDeflationary}{\intensity[]^{-}}
\newcommand{\intensityInflationary}{\intensity[]^{+}}
\newcommand{\Nminus}{\countingProc[]^{-}}
\newcommand{\Nplus}{\countingProc[]^{+}}

\newcommand{\stateVariable}{X}
\newcommand{\transProb}{\phi}

\newcommand{\directImpact}{\text{Dir}}
\newcommand{\indirectImpact}{\text{Indir}}

\newcommand{\initialTime}{t_{0}}
\newcommand{\terminationTime}{\tau}
\newcommand{\metaorderSize}{Q_{0}}

\newcommand{\price}{P}
\newcommand{\tickSizeOfLOB}{\psi}
\newcommand{\midprice}{\midPrice}
\newcommand{\midPrice}{\price^{m}}
\newcommand{\volume}{V}

\newcommand{\bestBidPrice}{\price^{b}}
\newcommand{\nthBestBidPrice}[1][1]{\price^{b,{#1}}}
\newcommand{\bestAskPrice}{\price^{a}}
\newcommand{\nthBestAskPrice}[1][1]{\price^{a,{#1}}}

\newcommand{\nthBestBidVolume}[1][1]{\volume^{b,{#1}}}

\newcommand{\nthBestAskVolume}[1][1]{\volume^{a,{#1}}}
\newcommand{\volumeImbalance}{I}

\newcommand{\LOBspread}{\phi}

\usepackage{lipsum}
\newcommand\blfootnote[1]{%
  \begingroup
  \renewcommand\thefootnote{}\footnote{#1}%
  \addtocounter{footnote}{-1}%
  \endgroup
}
\usepackage{hyperref}
\definecolor{urlcolor}{rgb}{0,.145,.698}
\definecolor{linkcolor}{rgb}{.71,0.21,0.01}
\definecolor{citecolor}{rgb}{.12,.54,.11}
\hypersetup{
breaklinks=true,
colorlinks=true,
urlcolor=urlcolor,
linkcolor=linkcolor,
citecolor=citecolor,
}

    \usepackage{listings}
    \definecolor{stringcolor}{rgb}{0.3,0.0,0.0}
    \definecolor{commentedcode}{rgb}{0.1,0.2,0.1}
    \definecolor{codegray}{rgb}{0.5,0.5,0.5}
    \definecolor{backcolour}{rgb}{0.965,0.965,0.965}
    \lstdefinestyle{mystyle}{
        backgroundcolor=\color{backcolour},   
        commentstyle=\color{commentedcode},
        numberstyle=\tiny\color{codegray},
        stringstyle=\color{stringcolor},
        basicstyle=\ttfamily\scriptsize,
        breakatwhitespace=false,         
        breaklines=true,                 
        captionpos=t,                    
        keepspaces=true,                 
        numbers=left,                    
        numbersep=5pt,                  
        showspaces=false,                
        showstringspaces=false,
        showtabs=false,                  
        tabsize=4
    }
    \newcommand{\codehl}[1]{\colorbox{gray!8}{\texttt{\small #1}}}

\title{Non-average price impact in order-driven markets}
\author{Claudio Bellani$^1$, Damiano Brigo$^1$,\\ Mikko S.~Pakkanen$^1$, and Leandro S\'anchez-Betancourt$^2$
\blfootnote{$^1$Dept. of Mathematics, Imperial College London. $^2$Dept. of Mathematics, King's College London. Correspondence email: leandro.sanchez-betancourt@kcl.ac.uk}
}
\date{\today}

\begin{document}
\maketitle
\begin{quotation}
\textbf{Abstract.}
We present a measurement of price impact in order-driven markets 
that does not require averages across executions or scenarios.
Given the order book data associated with one single execution of a sell metaorder,
we measure its contribution to price decrease 
during the trade. 
We do so by modelling the limit order book
using state-dependent Hawkes processes,
and by defining the price impact profile of the execution
as a function of the compensator of a stochastic process in our model. 
We apply our measurement to a dataset from NASDAQ,
and we conclude that the clustering of sell child orders
has a bigger impact on price than their sizes. 
\end{quotation}

\section{Introduction}
Price impact is the phenomenon whereby trade executions affect the price of the asset being traded. 
The price is affected in a way that is unfavourable to the trade, 
i.e. it decreases as a consequence of sell orders 
and it increases as a consequence of buy orders.  
Hence, price impact is often listed as a hidden transaction cost,
and price impact detection is considered a branch of transaction cost analysis. 

References in the literature are numerous and among them are 
\cite{Tor97bar}, 
\cite{ATHL05dir}, 
\cite{MVMGFVLM09mar}, 
\cite{TLDLKB11ano}, 
\cite{EFR12mea}, 
\cite{BILL15mar}, 
\cite{BSKB15slo}, 
\cite{ZTFL15bey}, 
\cite{TEB16squ}, and 
\cite{PB18uni}.
An overview on the investigations about price impact was provided by 
\cite{Bou17mar}. 

Recently, some authors (e.g., \cite{CC19vol}) 
pointed out that 
the findings about price impact laws 
overlap with 
the scaling of price variations with the square-root of time, 
namely the volatility,
and they go as far as questioning the existence of 
price impact altogether. 

However,
as observed in \cite{Bou17mar}, 
price impact can be regarded as an embodiment of basic economic principles about supply and demand, 
and it must play a role in price formation. 
This is to say that, 
despite the acknowledged overlap between execution-induced price moves and volatility-induced ones, 
the phenomenon of price impact must exist in a functioning market.
The consensus among these authors is that,
in order to detect it,
one has to average across several executions and scenarios to factor out volatility -- 
see \cite{BMBB19imp}.

In this paper, 
we present a measurement of price impact in order-driven markets 
that does not require averages across executions and scenarios.
Given the order book data associated with one single execution of a sell metaorder 
(subdivided into smaller child orders distributed in time), 
we measure its contribution to price decrease 
(or to the impediment of price increase) 
during the trade. 

Our motivation is twofold. 
On the one hand, 
the economic argument for which price impact must exist applies scenario-by-scenario and to every execution. 
Therefore, it is interesting to wonder whether one might refrain from taking averages and yet detect it. 
This means questioning whether the averaging operation is the only viable way to factor out volatility. 
On the other hand -- and on a less academic ground --, 
investors and traders might in fact be more interested in 
a trade-by-trade assessment of price impact 
rather than on emergent universal properties of price moves during executions. 
This was pointed out in the aforementioned article by \cite{CC19vol}, 
who argue that 
for institutional investors who execute infrequent but large trades, 
<<the observed impact may deviate significantly from such an average>>.

Our measurement of price impact is based on a granular model of order book dynamics. 
Our model extends that of 
\cite{BM14haw},
where  a four-dimensional Hawkes process was used to describe arrival of market orders into a limit order book and price changes. 
We increase the granularity by moving the description of price changes to a state variable that evolves in time, 
and by tracking not only market orders but also limit orders. 
The pair Hawkes process - state variable is modelled on a class of hybrid marked point processes introduced in 
\cite{MP20hyb}
and called \emph{state-dependent Hawkes processes}.

The so-increased granularity allows us to: 
(\emph{i}) have a snapshot-by-snapshot proxy for volumes of offers on every level of the order book;
(\emph{ii}) assess the impact that a labelled agent has on the market 
without assuming that their orders walk the book with the same frequency as other participants' orders do, 
as was instead assumed in \cite{BM14haw}. 

In an average measure of price impact, one fixes a trade direction for order executions and looks at price movements across several scenarios. 
Since in these scenarios the signs of all other market variables are on average expected to cancel out, 
an average price movement emerging during these executions represents the impact sought,
in that the trade direction was the only isolated variable.
Instead, in a non-average measure of price impact, 
one cannot count on this average cancellation of all market variables other than the trade direction of the considered execution. 
This poses the difficult hypothetical question of what would have happened if the trade under consideration had not been executed. 
In other words, the execution-induced price movement during the single observed scenario has to be filtered from the price movement induced by all the surrounding market noise. 
We do so by shaping our notion of non-average price impact around the concept of compensator of a counting process. 
This allows us to focus on a single realisation of the process and yet factor out its volatility.

We demonstrate our measurements by assessing price impact on NASDAQ data for the ticker INTC. 
Since market participants behind the observed market events are not identified, 
we cannot directly measure the impact of a chosen execution among those in the raw data. 
Instead, we calibrate our market model on the provided data, and then we simulate a liquidation in the calibrated model.
By so doing, we illustrate the usefulness of the class of market replayers to which our model belongs. 
Concretely, this usefulness is the possibility to answer `what-if' queries, namely the sort of hypothetical questions that make non-average price impact detection challenging.

More precisely, we fix the size of the metaorder to be executed, and we examine how the price impact of such an execution varies with (\emph{i}) the size of the child orders (whether they are likely to consume all available liquidity on the first price level of the limit order book or not); (\emph{ii}) the scheduling of the child orders (whether they cluster around particular times or are evenly distributed during the execution). We conclude that in the examined dataset, the clustering of child orders has a bigger impact on the price than their sizes.

The paper is organised as follows.
Section \ref{sec.order-driven_markets}
gives some background about order-driven markets
and limit order books.
Section \ref{sec.counting_processes}
recalls basic notions from the theory of counting processes
and establishes the notation used in the paper.
Section \ref{sec.sdhawkes_model}
describes our model for order-driven markets 
based on state-dependent Hawkes processes.
Section \ref{sec.price_impact_profiles}
defines price impact in our model,
and 
describes how we quantify it.
Section \ref{sec.applications}
presents applications of our modelling framework
to a dataset from NASDAQ,
and 
draws some conclusions on the main drivers of price impact in this dataset.
Finally, Appendix \ref{sec.proofs} collects the proofs of the propositions in the paper.

\section{Background on order-driven markets and limit order books}
\label{sec.order-driven_markets}
Order-driven markets are trading venues organised around \emph{limit orders}. 
A limit order is the fundamental action that market participants in order-driven markets can perform.
It is represented by a 4-tuple $(t,q,p,d)$, 
where $t$ denotes time, $q$ denotes size, $p$ denotes price,
and $d$ denotes direction. 
We say that a market participant \emph{posts} a limit order $(t,q,p,d)$ if at time $t$ they submit to the exchange their commitment to buy ($d=1$) or sell ($d=-1$) the amount $q$ at the price $p$. 
The  price $p$ is interpreted as the highest price at which they are committed to buy if $d=1$,
or the lowest price at which they are committed to sell if $d=-1$.
By regulation, the price $p$ must be an integer multiple of some fixed $\tickSizeOfLOB>0$ known as the tick size.
Market participants have also the possibility to withdraw their commitments to trade,
by \emph{cancelling} a previously submitted limit order. 

A trading epoch in an order-driven market is the collection of all the limit orders submitted by market participants within some time interval.
Limit orders cannot be submitted simultaneously,
so that a limit order is identified by its timestamp $t$.
We express this mathematically by saying that a trading epoch is the graph of a function from a subset of the positive half-line to the space $\lbrace (q,p,d): \,
q\geq0,
p\geq0,
d=-1,1\rbrace$. 

When a market participant submits a limit order $(t,q,p,d)$,
a matching algorithm run by the trading platform looks for possible counterparts to their order:
if other participants' commitments to trade exist to (partially) fulfill the order
(at a price not worse than $p$ from the submitter's point of view),
then their order is (partially) cleared.
The fraction of their order that is cleared disappears from the market,
and it is said to be \emph{executed};  
the fraction that could not be fulfilled is recorded in the \emph{limit order book},
waiting for counterparts to trade with.

A limit order $(t,q,p,d)$ in a trading epoch is said 
to be \emph{active} (or outstanding) at time $u$ 
if $t\leq u$ and by time $u$ the order has neither been executed nor been cancelled.  

The search for possible trading counterparts for the matching of the incoming order $(t,q,p,d)$ happens among the active orders with timestamp $s<t$ and opposite direction $-d$. 
We say that the active order $(s,\rho,\pi,-d)$ matches the incoming order  $(t,q,p,d)$
if 
$\pi d \leq p d$;
in this case, the two orders are executed one against the other,
and $\rho \wedge q$ units of shares are transacted at the price $\pi$ per unit of share. 
After the match, the matched orders are replaced with the orders $(s,\tilde{\rho}, \pi, -d)$ and $(t,\tilde{q}, p, d)$, 
where $\tilde{\rho} = \rho - \rho \wedge q$ 
and $\tilde{q} = q - \rho \wedge q$. 
At least one of them has null size.
If $\tilde{\rho} = 0$, i.e. $q \geq \rho$, the order with time stamp $s$ ceases to be active and it is deleted from its queue. 
If $\tilde{q} = 0$, i.e. $q\leq \rho$, then the order with time stamp $t$ (i.e. the incoming order) is fully executed,
and the search for possible matching counterparts stops. 
If instead $\tilde{q} > 0$, i.e. $q>\rho$, then the search continues among active orders with opposite direction $-d$. 

\begin{defi}\label{def.orderqueue}
 Let $E$ be a trading epoch in an order-driven market. 
	The ask order queue $A_t$ at time $t$ is defined as 
 \begin{equation*}
  A_t := \lbrace (s,q,p,-1)\in E : \, (s,q,p,-1) \text{ is active at time } t \rbrace. 
 \end{equation*}
 Similarly, the bid order queue $B_t$ at time $t$ is defined as 
 \begin{equation*}
  B_t := \lbrace (s,q,p,+1) \in E : \, (s,q,p,+1) \text{ is active at time } t \rbrace. 
 \end{equation*}
\end{defi}

A limit order book is a grid of equally spaced prices at which active limit orders sit. 
The space between consecutive grid nodes is called the tick size of the LOB, which we denote by $\tickSizeOfLOB$. 
All orders submitted to the exchange must have prices that are integer multiples of the tick size.
Prices are increasing from left to right.
At every node of the grid, outstanding limit orders to buy or sell at the corresponding price are collected;
these limit orders are also said to be queuing there.
Spontaneously -- i.e.
by market forces  -- such orders organise in a way that buy offers will be displaced on the left (the so-called \emph{bid side}) and sell offers will be on the right (the so-called \emph{ask side}).
Indeed, 
if there were a buy (resp., sell) limit order to the right (resp., left) of some sell (resp., buy) limit orders, or at the same node, 
then the matching algorithm would have matched them, 
clearing them out from the order book.  
Therefore, the whole configuration of a limit order book at time $t$ is given when the following variables are specified:
\begin{enumerate}[label={(\emph{\roman{*}})}, ref={\emph{\roman{*}}.}]
\item\label{item.bestaskprice} the best ask price $\bestAskPrice_{t}$, i.e. the lowest price at which one can find sell offers active at time $t$;
\item\label{item.bestbidprice} the best bid price $\bestBidPrice_t$, i.e. the highest price at which one can find buy offers active at time $t$;
\item\label{item.askvolume} the volume $\nthBestAskVolume[i]_t$ of ask offers at price $\nthBestAskPrice[i]_t= \bestAskPrice_t + (i-1)\tickSizeOfLOB$, for $i=1,2,\dots$, i.e. the quantity 
\begin{equation*}
 \nthBestAskVolume[i]_t = \sum_{\lbrace(s,q,\nthBestAskPrice[i]_t,-1) \in A_t : \quad s\leq t\rbrace} q,
\end{equation*}
where the sum is over all the active sell limit orders submitted by time $t$ for  a price of $\bestAskPrice_t + (i-1)\tickSizeOfLOB$;
\item\label{item.bidvolume} the volume $\nthBestBidVolume[i]_t$ of bid offers at price $\nthBestBidPrice[i]_t=\bestBidPrice_t - (i-1)\tickSizeOfLOB$, for $i=1,2,\dots$, i.e. the quantity 
\begin{equation*}
 \nthBestBidVolume[i]_t = \sum_{\lbrace (s,q,\nthBestBidPrice[i]_t,1) \in B_t: \quad s\leq t\rbrace} q,
\end{equation*}
where the sum is over all the active buy limit orders submitted by time $t$ for  a price of $\bestBidPrice_t - (i-1)\tickSizeOfLOB$.
\end{enumerate} 

The whole configuration of a limit order book at time $t$ is described by the variables $(\bestAskPrice_t, \bestBidPrice_t, \lbrace (\nthBestAskVolume[i]_t, \nthBestBidVolume[i]_t) : \, \,
i=1,2,\dots\rbrace)$.
Apart from these variables,
other derived quantities are useful to assess the properties of an order book.
These are the spread,
the mid-price and the volume imbalance (or queue imbalance). 

The spread at time $t$ is the distance $\lvert \bestAskPrice_t - \bestBidPrice_t \rvert$ between best ask price  and best bid price,
which we denote by $\LOBspread_t$.
The mid-price at time $t$ is the mid point in between  $\bestBidPrice_t$ and $\bestAskPrice_{t}$,
which we denote by $\price^{m}_t$,
and is given by $\price^m_t = (\bestAskPrice_t + \bestBidPrice_{t})/{2}$.
Finally,
the $n$-levels volume imbalance (or queue imbalance) at time $t$,
denoted $\volumeImbalance^{n}_t$,
is the normalised excess of 
limit orders on the first $n$ levels of the bid side 
compared to the limit orders on the first $n$ levels of the ask side,
namely
\begin{equation}\label{eq.definition_volume_imbalance}
\volumeImbalance^{n}_t =
\frac{\sum_{i\leq n}\nthBestBidVolume[i]_t - \sum_{i\leq n}\nthBestAskVolume[i]_t}{\sum_{i\leq n}\nthBestBidVolume[i]_t + \sum_{i\leq n}\nthBestAskVolume[i]_t},
\end{equation}
where $\sum_{i\leq n}\nthBestBidVolume[i]_t$ (resp., $\sum_{i\leq n}\nthBestAskVolume[i]_t$) is the cumulative volumes on the first $n$ bid (resp., ask) levels. 
The queue imbalance $\volumeImbalance^n$ is widely accepted as a reliable signal for the next mid-price move (see \cite{CDJ18enh}):
when it is close to -1 the mid-price will likely decrease,
and when it is close to +1 it will likely increase. 

The arrival of a limit order to the market triggers two events:
one is the consumption of the liquidity capable to (partially) service the limit order,
the other is the addition of the non-executed part of the limit order to the appropriate queue in the order book.
Proposition \ref{prop.decomposition_of_limit_order} states the terms of this decomposition.

\begin{prop}\label{prop.decomposition_of_limit_order}
 Given the configuration $(\bestAskPrice_{t-}, \bestBidPrice_{t-},\lbrace  (\nthBestAskVolume[i]_{t-}, \nthBestBidVolume[i]_{t-}) : \, \, i=1,2,\dots\rbrace )$ of the limit order book immediately before time $t$, processing the sell limit order $(t,q,p,-1)$ is equivalent to processing the pair of orders $[(t,q^-_M,0,-1),(t,q-q^-_M,p,-1) ]$, with the former having priority over the latter, and where 
 \begin{equation*}
  q^-_M:= \min\left(q, \sum_{i\geq 1} \nthBestBidVolume[i]_{t-} \one \left\lbrace \nthBestBidPrice[i]_{t-}\geq p\right\rbrace\right).
 \end{equation*}
Similarly, processing the buy limit order $(t,q,p,+1)$ is equivalent to processing the pair of orders $[(t,q^+_M,\infty,+1),(t,q-q^+_M,p,+1) ]$, where 
 \begin{equation*}
  q^+_M:= \min\left (q, \sum_{i\geq 1} \nthBestAskVolume[i]_{t-} \one \left\lbrace \nthBestAskPrice[i]_{t-}\leq  p\right\rbrace\right).
 \end{equation*}
\end{prop}

From this point forward, given a limit order $(t,q,p,d)$, we let $q_M=q^+_M$ if $d=+1$, and $q_M=q^-_M$ if $d=-1$.
The first component $(t,q_M,0,-1)$ in the decomposition $[(t,q_M,0,-1),(t,q-q_M,p,-1) ]$ of the sell limit order $(t,q,p,-1)$ is called \emph{sell market order}.
Notice that in the 4-tuple $(t,q_M,0,-1)$,
the price $p$ is set to zero,
and this guarantees immediate execution:
the sale of the amount $q_M$ is instantaneously matched with outstanding buy limit orders on the bid side,
and no fraction of $q_M$ is put into the queue.
On the contrary,
the second component $(t,q-q_M,p,-1)$ specifies exactly that part of $(t,q,p,-1)$ which will be queued. 
Similarly, $(t,q_M,\infty,+1)$ represents the market order component of the buy limit order $(t,q,p,+1)$,
and it is referred to as \emph{buy market order}.
The price $p=\infty$ is a way to express the fact that a counterpart for the purchase of the amount $q_M$ will instantaneously be found on the ask side of the order book.
In the following,
the term 'market order' (either buy or sell) will be referring to the first component  in the decomposition of a limit order with non-zero market order size $q_M >0$.\footnote{
In some trading venues,
traders can actually submit market orders,
i.e.
buy or sell orders that are executed without price constraints,
at least as long as offers with the opposite direction exist.
Even in such cases,
we keep our convention of referring to the fraction of a limit order that is executed upon submission as market order.}  

A sell market order $(t,q_M,0,-1)$ is said to `walk the book' if $q_M > \nthBestBidVolume[1]_t$.
Similarly,
a buy market order walks the book if its size is larger than the volume of offers on the first ask level. 

Given a time window
$\timeWindow$,
the evolution in time of the limit order book
$\lbrace (\bestAskPrice_t, \bestBidPrice_t,\lbrace (\nthBestAskVolume[i]_t, \nthBestBidVolume[i]_t) : \, \,
i=1,2,\dots\rbrace):  \quad 0\leq t \leq \timeHorizon \rbrace$
results from the history of all limit order submissions,  cancellations and executions that happened in
$\timeWindow$.
If every limit order in  
$\lbrace (t,q,p,d): \,
0\leq t\leq \timeHorizon\rbrace$
is decomposed as per in Proposition \ref{prop.decomposition_of_limit_order},
then the seller-initiated trades that happened in
$\timeWindow$
are
$\lbrace (t,q_M,0,-1):\, \,  q_M>0, \,
0\leq t\leq \timeHorizon \rbrace$,
and the buyer-initiated trades that happened in
$\timeWindow$
are
$\lbrace (t,q_M,\infty,+1):\, \,  q_M>0, \,
0\leq t\leq \timeHorizon \rbrace$.  
Hence,
in the following we will identify trades with market orders of non-zero size $q_M>0$.

\section{Background on counting processes and Hawkes processes}
\label{sec.counting_processes}
\subsection{Counting processes}
In this section we introduce our notation for counting processes,
and we review basic concepts from the theory of such processes.
Our main reference is \citealp[Chapter 14]{DVJ08int}.

Let $\numEventTypes$ be a positive integer. For each $e$ ranging from $1$ to $\numEventTypes$, let $\arrivalTimes_{j}$, $j=1, 2, \dots$, be a strictly increasing sequence of positive random times, and assume that $\arrivalTimes_{j}\neq\arrivalTimes[e\derivative]_{j\derivative}$ if $(e,j)\neq (e\derivative,j\derivative)$. Then, 
\begin{equation*}
 \countingProc(t):= \sum_{j} \one\left\lbrace\arrivalTimes_{j}\leq t \right\rbrace, \qquad t\geq 0,
\end{equation*}
is a non-decreasing right-continuous process; we call $\countingProc$ the counting process associated with the sequence $(\arrivalTimes_{j})_j$. Notice that $(\arrivalTimes_{j})_j$ can be retrieved from $\countingProc$ by 
\begin{equation*}
 \arrivalTimes_{j} = \inf \left\lbrace t>0: \, \countingProc(t)\geq j\right\rbrace;
\end{equation*}
hence, there is a one-to-one correspondence between $\countingProc$ and $(\arrivalTimes_{j})_j$. 

For $t>0$ we define 
\begin{equation*}
 \delta\countingProc(t):= \lim_{h\downarrow 0} \Big( \countingProc(t) - \countingProc(t-h) \Big),
\end{equation*}
and we notice that $\delta \countingProc(t) = 1$ if and only if $t=\arrivalTimes_{j}$ for some $j$, otherwise $\delta \countingProc(t) = 0$.

The $\numEventTypes$-dimensional vector $\multiCountingProc(t) = (\countingProc[1](t), \dots , \countingProc[\numEventTypes](t))$ is referred to as multivariate counting process associated with the $\numEventTypes$ sequences $(\arrivalTimes_{j})_j$, $e=1,\dots, \numEventTypes$. Let $\groundProc(t) := \countingProc[1](t)+ \dots + \countingProc[\numEventTypes](t)$ be the ground process of $\multiCountingProc$, and let 
\begin{equation*}
 \arrivalTimes[ ]_{n} := \inf \left\lbrace t>0: \, \groundProc(t) \geq n\right\rbrace, \qquad n=1,2,\dots, 
\end{equation*}
be the ordered sequence of random times stemming from the union $\lbrace \arrivalTimes_{j}: \, j=1,2,\dots; \,\, e=1,\dots,\numEventTypes\rbrace$. By defining for $n=1,2, \dots$, 
\begin{equation*}
 \event := \sum_{e=1}^{\numEventTypes} e \, \one\left\lbrace \delta \groundProc(\arrivalTimes[ ]_{n}) = \delta\countingProc(\arrivalTimes[ ]_{n}) \right\rbrace,
\end{equation*}
we have that the pair $(\arrivalTimes[ ]_{n}, \event[n])$ equivalently characterises the multivariate counting process, because 
\begin{equation}\label{eq.NTE_counting_proc}
 \countingProc(t) = \sum_{n} \one \left\lbrace \arrivalTimes[ ]_{n}\leq t, \, \event[n] = e\right\rbrace,
\end{equation}
for all $t>0$ and all $e=1,\dots, \numEventTypes$.

We can interpret this construction by saying that the index $e$ labels $\numEventTypes$ types of events that occur in time, and $\countingProc(t)$ counts the number of events of type $e$ that have occurred by time $t$. 

\begin{example}[``Poisson process'']
 Let $\tau_e^j$, $j=1,2,\dots$, $\, \, e=1,\dots, \numEventTypes$ be independent random variables such that $\tau_e^j$ is exponentially distributed with parameter $\intensity>0$, $\, e=1,\dots, \numEventTypes$. Let $\arrivalTimes_{j} := \sum_{k\leq j} \tau_e^{k}$, and notice that $\arrivalTimes_{j}$ has probability density function 
 \begin{equation*}
  f_{e,j}(t) = \frac{\intensity^j}{(j-1)!}t^{j-1} e^{-\intensity t} \one\lbrace t>0\rbrace.
 \end{equation*}
 Then, the multivariate counting process $\multiCountingProc$ associated with the arrival times $\arrivalTimes_{j}$ is called $\numEventTypes$-dimensional Poisson process of rates $\intensity[1], \dots \intensity[\numEventTypes]$. This name is justified as follows. 
 Since $\lbrace \countingProc(t)\geq j\rbrace = \lbrace \arrivalTimes_{j} \leq t\rbrace$, we have that $\frac{d}{dt} \Prob(\countingProc(t) \geq j) = f_{e,j} (t)$. On the other hand, if we define 
 \begin{equation*}
  S_{e,j}(t) := \sum_{k\geq j} \frac{(\intensity t)^k}{k!}e^{-\intensity t},
 \end{equation*}
we also have that $\frac{d}{dt} S_{e,j}(t) = f_{e,j}(t)$, by telescopic sum. Since $S_{e,j}(0) = \Prob (\countingProc(t)\geq 0)$, we deduce that $\Prob(\countingProc(t) \geq j) = S_{e,j}(t)$, and that
\begin{equation*}
 \Prob\left( \countingProc(t) = j\right) = 
 \frac{(\intensity t)^j}{j!} \exp\left( -\intensity t\right).
\end{equation*}
Therefore for every $t$, $\countingProc(t)$ is a Poisson random variable of parameter $\intensity t$, and the ground process $\groundProc$ of $\multiCountingProc$ is such that for every $t$, $\groundProc(t) \sim \text{Pois}(\intensity[1]t + \dots + \intensity[\numEventTypes]t)$.
\end{example}

The minimal filtration to which a multivariate counting process $\multiCountingProc$ is adapted -- and such that it satisfies the usual conditions of completeness and right-continuity -- is called the internal history of $\multiCountingProc$. Any other filtration to which $\multiCountingProc$ is adapted is called a history of $\multiCountingProc$, and it must be a superset of the internal history. 

\begin{defi}\label{def.compensator}
 Let $\stochasticBase$ be a filtered probability space where the multivariate counting process $\multiCountingProc$ is defined, and assume that $\filtrationF$ is a history of $\multiCountingProc$. We say that the $\numEventTypes$-dimensional stochastic process $\compensator = (\compensator_1, \dots , \compensator_{\numEventTypes})$ is an $\filtrationF$-compensator for $\multiCountingProc$ if:
 (\emph{i}) $\compensator(0)=0$ and $\compensator$ is of finite variation;
 (\emph{ii}) $\compensator$ is $\filtrationF$-predictable;
 (\emph{iii}) $\compensator$ is right-continuous;
 (\emph{iv}) $\multiCountingProc - \compensator$ is a local martingale. 
\end{defi}


Given the counting process $\multiCountingProc$ and a history $\filtrationF$, the $\filtrationF$-compensator is unique up to an evanescent set, and it is equivalently characterised as the $\filtrationF$-predictable projection of $\multiCountingProc$, namely as the $\filtrationF$-predictable non-decreasing process $\compensator$ such that 
\begin{equation}\label{eq.compensators_and_predict_projections}
 \Expectation \left[ \int_{\R_+} Y d\multiCountingProc\right]
 = 
 \Expectation \left[ \int_{\R_+} Y d\compensator\right]
\end{equation}
for all non-negative $\filtrationF$-predictable processes $Y$ (see \citealp[Proposition 14.2.II]{DVJ08int}). 

If $\compensator$ is absolutely continuous, we write 
\begin{equation*}
 \compensator (t) = \intzerot \intensity[ ](s)ds,
\end{equation*}
for some $\filtrationF$-predictable process $\intensity[ ]=(\intensity[1],\dots,\intensity[\numEventTypes])$,
which is called intensity of the counting process $\multiCountingProc$. 
Combining this with equation \eqref{eq.compensators_and_predict_projections}, 
one obtains the formula 
\begin{equation*}
 \Expectation\left[ \countingProc(t)-\countingProc(s) \vert \filtrationF_s \right]
 =
 \Expectation\left[ \int_{s}^{t} \intensity(u) du
  \vert \filtrationF_s \right],
  \qquad  s\leq t,
\end{equation*}
which allows to interpret $\intensity(t)$ as a measure of the ``instantaneous risk'' of a jump at time $t$ in the $e$-th component of the counting process $\multiCountingProc$. Notice that this ``risk'' evolves in time and it varies depending on the information available up to time $s$. 

Compensators are crucial in the following time-change result,
which will be used to perform goodness-of-fit diagnostics 
(see Section \ref{sec.calibration}).

\begin{thm}[{\cite{Mey71dem}}]\label{thm.meyer1971}
Let $\multiCountingProc$ be a $\numEventTypes$-dimensional counting process with arrival times $\arrivalTimes$. Assume that $\multiCountingProc$ has continuous compensator $\compensator$ such that $\compensator_{e}(t) \rightarrow \infty$ as $t\rightarrow \infty$ for all $e=1, \dots, \numEventTypes$. Then, the random sequences $\lbrace \compensator(\arrivalTimes_{j}): \, j=1,2,\dots \rbrace$, $e=1,\dots, \numEventTypes$ are the arrival times of a $\numEventTypes$-dimensional unit-rate Poisson process, namely the time-changed inter-arrival times
\begin{equation}\label{eq.time-changed_intertimes}
 \tau_e^j := \compensator(\arrivalTimes_{j}) - \compensator(\arrivalTimes_{j-1})
\end{equation}
are all independent exponentially distributed random variables for $j=1,2,\dots$ and $e=1,\dots,\numEventTypes$.  
\end{thm}
For a proof of Theorem \ref{thm.meyer1971}, see \cite{BN88sim}. 

\subsection{Multidimensional Hawkes processes}
In this section, we collect some elements of the theory of state-dependent Hawkes processes from 
\cite{MP18sta}
and
\cite{MP20hyb}.

\begin{defi}\label{def.hawkesprocess}
A $\numEventTypes$-dimensional counting process $\multiCountingProc$ is called Hawkes process if it admits an absolutely  continuous compensator $\compensator$ with intensities 
\begin{equation}\label{eq.intensity_ordHawkes}
 \intensity(t) = \baseRate_e + \sum_{\eone=1}^{\numEventTypes}\intzerot \hawkesKernel\subscriptee(t-s)d\countingProc[\eone](s), \qquad e=1,\dots,\numEventTypes,
\end{equation}
for some non-negative base rates $\baseRate_e \geq 0$, and some non-negative locally integrable functions $\hawkesKernel\subscriptee \geq 0$ that are supported on the non-negative half line.
\end{defi}

The matrix-valued function $t\mapsto [\hawkesKernel\subscriptee(t)]_{e,\eone = 1, \dots, \numEventTypes}$ is referred to as the kernel of the Hawkes process $\multiCountingProc$.
 If all the kernel functions are integrable, the spectral radius $\rho$ of the $\numEventTypes\times\numEventTypes$-matrix of $\Lone$ norms $\Vert \hawkesKernel\subscriptee \Vert_{1}$ is called radius of the Hawkes kernel;
if some of the kernel functions are not integrable, the spectral radius is set to $+\infty$.

A $\numEventTypes$-dimensional Hawkes process is asymptotically stationary 
if the radius of its kernel is smaller than $1$; in this case the intensity process $\intensity[ ]$ is asymptotically stationary.

Let $\stateSpace$ be a finite state space.
We can label its elements as $x=1, \dots, \numStates$, where $\numStates$ is the number of possible states of the system.
A state-dependent counting process is a pair $(\multiCountingProc, \stateVariable)$, 
where for all $t$, $\multiCountingProc(t)$ records the number of events occurred by time $t$ as per formula \eqref{eq.NTE_counting_proc}, 
and $\stateVariable(t)$ records the state of the system at time $t$.
More specifically, we have:
\begin{defi}[{\citealp[Definition 2.1]{MP18sta}}]\label{def.sdHawkes}
 Let $\multiCountingProc$ be a $\numEventTypes$-dimensional counting process.
Let $\stateVariable$ be a continuous-time piecewise-constant process in the finite state space $\stateSpace$ of cardinality $\numStates$.
Let $\internalHistory$ be the minimal complete right-continuous filtration generated by the pair $(\multiCountingProc, \stateVariable)$.
Then, we say that $(\multiCountingProc, \stateVariable)$ is a state-dependent Hawkes process if 
 \begin{enumerate}[label={(\emph{\roman{*}})}]
  \item $\multiCountingProc$ admits an absolutely continuous $\internalHistory$-compensator with intensities
  \begin{equation}\label{eq.intensity_of_sdHawkes}
   \intensity(t) = \baseRate_e + \sum_{\eone=1}^{\numEventTypes}\int_{[0,t)} \hawkesKernel\subscriptee(t-s, \stateVariable(s))d\countingProc[\eone](s), \qquad e=1,\dots,\numEventTypes,
  \end{equation}
  for some $\numEventTypes$ non-negative base rates $\baseRate_e\geq 0$, $\, e=1, \dots, \numEventTypes$, 
		 and 
		 some $\numEventTypes\squared$ measurable functions $\hawkesKernel\subscriptee: \R_+ \times \stateSpace \rightarrow \R_+$, $e, \eone = 1, \dots, \numEventTypes$,  such that $\hawkesKernel\subscriptee(\cdot, x)$ is locally integrable for all $x$ in $\stateSpace$;
  \item $\stateVariable$ jumps only at arrival times $\arrivalTimes[ ]_{n}$ of $\multiCountingProc$, and there exist $\numEventTypes$ transition matrices $\transProb_e(\cdot,\cdot)$, $e=1,\dots, \numEventTypes$, defined on $\stateSpace$ such that for all $n$
  \begin{equation}\label{eq.markov_update_stateVariable}
   \Prob\left( \stateVariable(\arrivalTimes[ ]_{n})=x \,  \vert \,  E_n, \, \internalHistory_{\arrivalTimes[ ]_{n}-} \right)
   =
   \transProb_{E_n}\left(\stateVariable(\arrivalTimes[ ]_{n}-),x\right),
   \qquad
   x=1, \dots, \numStates,
  \end{equation}
  where $\stateVariable(\arrivalTimes[ ]_{n}- ) = \lim_{t\uparrow \arrivalTimes[ ]_{n}} \stateVariable(t)$ is the state of the system immediately before the $n$-th event $E_n$, and $\internalHistory_{\arrivalTimes[ ]_{n}-} = \bigvee_{\epsilon>0} \internalHistory_{\arrivalTimes[ ]_{n} - \epsilon}$ represents the information available immediately before this event.
 \end{enumerate}
\end{defi}

Given a state-dependent Hawkes process $(\multiCountingProc, \stateVariable)$, 
let $\arrivalTimes[]_n$ and $\event[n]$ be the sequences of arrival times and events that equivalently describe the counting process component $\multiCountingProc$ of the pair $(\multiCountingProc, \stateVariable)$, as per equation \eqref{eq.NTE_counting_proc}.
Let $\stateVariable_n$ be the sequence of states $X(\arrivalTimes[]_n)$, for $n=1,2,\dots$.
Then, the 
$\numEventTypes\numStates$-dimensional counting process
\begin{equation}\label{eq.hybridHawkes}
 \hybridHawkes_{e,x}(t) := 
 \sum_{n} \one \left\lbrace \arrivalTimes[]_n \leq t, \, \event[n] = e, \, 
 \stateVariable_n = x \right\rbrace
\end{equation}
is called the hybrid-MPP counterpart of  $(\multiCountingProc, \stateVariable)$.
We have that the $j$-th jump time $\arrivalTimes[e]_j$ of
the $e$-th component of
$\multiCountingProc$ is
the $j$-th order statistic of
$\lbrace \arrivalTimes[e,x]_k: \, k=1,2,\dots; \, \, x=1,\dots,\numStates \rbrace$,
where $(\arrivalTimes[e,x]_k)_k$ are the jump times of
the $(e,x)$-th component of
$\hybridHawkes$. 
Similarly, $\arrivalTimes[]_n$ is
the $n$-th order statistics of
$\lbrace \arrivalTimes[e,x]_k: \, k=1,2,\dots; \, \, e=1,\dots, \numEventTypes; \, \,  x=1,\dots,\numStates \rbrace$.
The $(e,x)$-th component $ \hybridHawkes_{e,x}$ of
the hybrid-MPP counterpart of
$(\multiCountingProc, \stateVariable)$ admits a continuous compensator with density given by 
\begin{equation}\label{eq.intensity_of_hybridMPP}
 \hybridIntensity_{e,x}(t) 
 =
 \transProb_{e}\left(\stateVariable(t),x\right)
 \left(
 \baseRate_e
 +\sum_{\eone,\, \xone} \int_{[0,t)} \hawkesKernel\subscriptee(t-s,\xone)d\hybridHawkes_{\eone,\xone}(s)
 \right),
\end{equation}
where $\transProb_{e}$ is the transition matrix associated with event type $e$, and $\hawkesKernel\subscriptee$, for $e,\, \, \eone = 1,\dots,\numEventTypes$, are the Hawkes kernels of $\multiCountingProc$.

Let $\bar{\intensity[]} = \intensity[1] + \dots + \intensity[\numEventTypes]$ be the sum of the intensities.
If $\bar{\intensity[]}$ is decreasing in time, then a state-depended Hawkes process $(\arrivalTimes[]_n, \event[n], \stateVariable_n)$ can be simulated as detailed in Algorithm \ref{algo.MP18_ogata}.

\begin{algorithm}
 \caption{{\citealp[Algorithm 2.4]{MP18sta}}}
 \label{algo.MP18_ogata}
 \begin{algorithmic}[5]
  \REQUIRE $(\arrivalTimes[]_i, \event[i], \stateVariable_i)_{i=1,\dots,n-1}$
  \STATE set $t:=\arrivalTimes[]_{n-1}$
  \STATE set $\xi:=0$
  \WHILE{$\xi=0$}
  \STATE draw $U \sim \text{Exp}(\bar{\intensity[]}(t))$
  \STATE set $\xi :=1$ with probability $\bar{\intensity[]}(t+U)/\bar{\intensity[]}(t)$
  \STATE update $t\leftarrow t+U$
  \ENDWHILE
  \STATE set $\arrivalTimes[]_n:=t$
  \STATE draw $\event[n]$ in $\lbrace 1,\dots,\numEventTypes\rbrace$ with probabilities proportional to $\lbrace \intensity[1](\arrivalTimes[]_n), \dots, \intensity[\numEventTypes](\arrivalTimes[]_n)\rbrace$
  \STATE draw $\stateVariable_n$ in $\lbrace 1,\dots,\numStates\rbrace$ with probabilities $\lbrace \transProb_{\event}(\stateVariable_{n-1},1), \dots,\transProb_{\event}(\stateVariable_{n-1},\numStates)\rbrace$
  \RETURN $(\arrivalTimes[]_n,\event[n],\stateVariable_n)$
 \end{algorithmic}
\end{algorithm}

\section{State-dependent Hawkes model}\label{sec.sdhawkes_model}
We consider four streams of random times: 
the stream $(\arrivalTimes[1]_j)_j$ of times when limit orders are executed on the bid side 
(equivalently identified with the arrival times of sell market orders);
 the stream  $(\arrivalTimes[2]_j)_j$ of times when limit orders are executed on the ask side 
(equivalently identified with arrival times of buy market orders); 
the stream $(\arrivalTimes[3]_j)_j$ of times when 
either an ask limit order is inserted inside the spread,
or the cancellation of a bid limit order depletes the liquidity available at the first bid level;
the stream $(\arrivalTimes[4]_j)_j$ of times when 
either a bid limit order is inserted inside the spread,
or the cancellation of an ask limit order depletes the liquidity available at the first ask level.

The four sequences of random times give rise to a four-dimensional counting process 
$\multiCountingProc=(\countingProc[1], \dots, \countingProc[4])$ 
with the following interpretation of its components:
\begin{itemize}
 \item $\countingProc[1](t)$ denotes the number of seller-initiated trades that happened before or at time $t$
(identified with the number of market orders arrived on the bid side of the order book by time $t$);
 \item $\countingProc[2](t)$ denotes the number of buyer-initiated trades that happened before or at time $t$
(identified with the number of market orders arrived on the ask side of the order book by time $t$);
 \item $\countingProc[3](t)$ denotes the number of decreases in the mid-price 
	 caused by a limit order insertion or cancellation that happened before or at time $t$;
 \item $\countingProc[4](t)$ denotes the number of increases in the mid-price 
	 caused by a limit order insertion or cancellation that happened before or at time $t$;
\end{itemize}

The counting process $\multiCountingProc$ is paired with the state variable $\stateVariable$. 
At time $t$, the state variable $\stateVariable(t)$ summarises the configuration 
$(\bestBidPrice_t,\bestAskPrice_t,\lbrace( \nthBestBidVolume[i]_t, \nthBestAskVolume[i]_t): \, \, i=1,2,\dots \rbrace)$ 
of the limit order book at time $t$, 
by recording a proxy for the $n$-levels volume imbalance, 
and the variation of the mid-price compared to time $t-$. 
More precisely, 
\begin{equation}\label{eq.state-variable}
 \stateVariable(t)
 =
 \begin{pmatrix}
 \stateVariable_1 (t)
 \\
 \stateVariable_2(t)
 \end{pmatrix}
 :=
 \begin{pmatrix}
 \one\lbrace\delta\midprice(t)>0\rbrace - \one\lbrace\delta\midprice(t)<0\rbrace
 \\
 \half\sum_{k=0}^{K-1} (2k-K+1 )\one\left\lbrace \frac{k-K}{K} \leq \volumeImbalance^n_t < \frac{2(k+1)-K}{K} \right\rbrace
 \end{pmatrix},
\end{equation}
where 
$\delta\midprice(t) = \lim_{\epsilon\downarrow 0} (\midprice(t) - \midprice(t-\epsilon) )$, 
and $\volumeImbalance^{n}_t$ was defined in equation \eqref{eq.definition_volume_imbalance}.
The first component $\stateVariable_1$ of the state variable $\stateVariable$ can take the values $-1$, $0$, $+1$, respectively denoting downward jump in the mid-price, unchanged mid-price, and upward jump in the mid-price.
The second component $\stateVariable_2$ of the state variable $\stateVariable$ 
is a discretisation of the $n$-levels queue imbalance $\volumeImbalance^{n}_t$, 
and -- assuming that $K$ is odd -- it takes integer values from $-(K-1)/2$ to $(K-1)/2$, 
spanning the full range of possible values of $\volumeImbalance^n$ from $-1$ to $+1$.

It follows from the definition of $\stateVariable_2$ that
 if at time $t$ we have that $\stateVariable_2 (t) = x_2$, 
 then the $n$-levels queue imbalance $\volumeImbalance^n_t$ at time $t$ must be in the half-open interval $[(2x_2 -K -1)/2K, \, \, (2x_2 +1)/K \, [$.
Notice that $\stateVariable_2$ depends on the two additional parameters $n$ and $K$: 
the former is the number $n$ of levels of the limit order books taken into account in the computation of the queue imbalance $\volumeImbalance^n$;
the latter is the number $K$ of points in the partition of the interval $[-1,1]$  used for the discretisation of $\volumeImbalance^n$.

The pair $(\multiCountingProc, \stateVariable)$ is modelled as a state-dependent Hawkes process,
hence we assume that there are 
base rates $\baseRate_e$, 
Hawkes kernels $\hawkesKernel\subscriptee=\hawkesKernel\subscriptee(t, \xone)$ 
and transition matrices $\transProb_e$ such that  Definition \ref{def.sdHawkes} is satisfied.
The number of event types is $\numEventTypes = 4$ 
and the number of states is $\numStates = 3K$.

When a new event occurs, 
i.e., when one of the components $\countingProc$ of $\multiCountingProc$ jumps, 
the state variable $\stateVariable$ is updated as per in equation \eqref{eq.markov_update_stateVariable}. 
The update models the mechanism whereby trades on either side of the limit order book can trigger 
changes in the mid-price and in the queue imbalance.
Indeed, 
assume that a sell (resp., buy) market order arrives at time $\arrivalTimes[1]_j$ (resp., $\arrivalTimes[2]_j$), 
and that $\stateVariable(\arrivalTimes[1]_j -) = (x_1,x_2)$ 
(resp.,  $\stateVariable(\arrivalTimes[2]_j -) = (x_1,x_2)$) 
for some $x_1$ in $\lbrace -1, 0, +1\rbrace$ and some $x_2$ in $\lbrace (1-K)/2, (3-K)/2, \dots, (K-1)/2 \rbrace$. 
Then, the mid-price jumps downward (resp., upward) with probability 
$p_{-}:=\sum_{y_2 = (1-K)/2}^{(K-1)/2} \transProb_{1}((x_1, x_2), (-1,y_2))$ 
(resp.,  $p_{+}:=\sum_{y_2 = (1-K)/2}^{(K-1)/2} \transProb_{2}((x_1, x_2), (+1,y_2))$), 
and it remains unchanged with probability 
$p_{0}:=1-p_{-}=\sum_{y_2 = (1-K)/2}^{(K-1)/2} \transProb_{1}((x_1, x_2), (0,y_2))$ 
(resp.,  $p_{0}:=1-p_{+}=\sum_{y_2 = (1-K)/2}^{(K-1)/2} \transProb_{2}((x_1, x_2), (0,y_2))$).\footnote{ There is no chance that a sell (buy) market order can cause an increase (decrease, respectively) in the mid-price.} 
This jump of the state variable happens exactly at the arrival time $\arrivalTimes[1]_j$ 
(resp.,  $\arrivalTimes[2]_j$)  of the sell (resp., buy) market order, 
and it naturally captures the mechanism responsible for the market-order-induced price change:
$p_{-}$ (resp.,  $p_{+}$) represents the probability that a sell (resp., buy) market order 
walks the book given its submission, 
and $p_{0}$  represents the probability that it does not. 
Notice that 
$p_{-}$ (resp.,  $p_{+}$) and $p_{0}$ 
depend on the state of the limit order book immediately before the arrival of the sell (resp., buy) market order, 
and in particular they depend on $x_2$. 
This is a granular description of the order book mechanism, 
and it accounts for the fact that 
it is less likely that a sell (resp., buy) market order walks the book 
when the volumes on the bid (resp., ask) side are high, 
namely 
$p_{-}(x_1,x_2) \leq p_{-}(x_1,\tilde{x}_2)$ if $x_2\geq \tilde{x}_2$ 
(resp.,  $p_{+}(x_1,x_2) \leq p_{+}(x_1,\tilde{x}_2)$ if $x_2\leq \tilde{x}_2$).

The first component $\stateVariable_1$ of the state variable $\stateVariable$ enables to 
write the following proxy for the mid-price:
\begin{equation}\label{eq.midprice_proxy}
 \midprice_0 + \frac{\tickSizeOfLOB}{2} \intzerot \stateVariable_1(s) d\groundProc(s),
\end{equation}
where $\tickSizeOfLOB$ is the tick size of the limit order book
and $\groundProc = \countingProc[1] + \dots +  \countingProc[4]$ 
is the ground process of $\multiCountingProc$. 

\begin{remark}\label{remark.BM14_model_comparison}
	Our model can be compared to that of \cite{BM14haw}. 
	In their model, 
	four streams of random times are considered:
	the stream $(\arrivalTimes[1]_j)_j$ of times when limit orders are executed on the bid side
	(equivalently identified with the arrival times of sell market orders);
	the stream  $(\arrivalTimes[2]_j)_j$ of times when limit orders are executed on the ask side
	(equivalently identified with arrival times of buy market orders);
	the stream  $(\arrivalTimes[3]_j)_j$ of times when the mid-price decreases;
	the stream  $(\arrivalTimes[4]_j)_j$ of times when the mid-price increases. 
	$\arrivalTimes[1]$ and $\arrivalTimes[2]$ are as in our model, whereas 
	$\arrivalTimes[3]$ and $\arrivalTimes[4]$ represent what in our model we represent through the state variable $\stateVariable_1$.
	In \cite{BM14haw},
	the four-dimensional counting process $\countingProc[] = \countingProc[](t)$ associated with
	$\arrivalTimes[1]$, $\arrivalTimes[2]$, $\arrivalTimes[3]$  and $\arrivalTimes[4]$ 
	is assumed to be a four-dimensional ordinary Hawkes process. 
	In their model, 
	a buy (resp., sell) market order coming into the exchange and walking the book at time $t$
	would be represented by the equation
	 $\delta\countingProc[2](t) = \delta\countingProc[4](t) = 1$ 
	(resp., $\delta\countingProc[1](t) = \delta\countingProc[3](t) = 1$).
	However, the components of a multidimensional Hawkes process jump simultaneously with probability zero.
	In other words, 
	if $\lbrace (\arrivalTimes[e]_j)_j: \, e=1,.\dots,4\rbrace$ are the arrival times associated with a $4$-dimensional Hawkes process, 
	it holds 
	$\Prob(\arrivalTimes[e]_j = \arrivalTimes[\eone]_{j\derivative}, \, \text{for some } j,j\derivative\geq 1, \text{ and } e\ne\eone) = 0$.
	Since the direction of the causality is unambiguous (a market order originates first and as a result of its execution the mid-price jumps), 
	\cite{BM14haw} propose to add to the Hawkes kernels $\hawkesKernel_{1,3}$ and $\hawkesKernel_{2,4}$ an atomic component. 
	This is the defining feature of the
	``impulsive impact kernel''
	-- see \cite[Section 2.1.3]{BM14haw}.
	In our paper, the usage of the state variable $\stateVariable_1$ circumvents the need of these atomic components
        and naturally accommodates mid-price changes triggered by market orders walking the book.
\end{remark}

The second component $\stateVariable_2$ of the state variable $\stateVariable$ reproduces 
the state variable of the queue-imbalance model in \cite{MP18sta}. 
It is conceived as the main indicator of 
the regime in which limit and market orders will arrive to the exchange: 
in high-frequency markets trading algorithms send their orders in response to 
observable quantities of the limit order book configuration, 
and a prominent one is indeed the queue imbalance. 
It is therefore expected that when $\stateVariable_2$ is positive (resp., negative), the intensities of events of types $e=2$ (resp.,  $e=1$) will be higher, 
because market participants following the queue imbalance signal will expect the price to increase (resp., decrease). 
After the price change,  
the volumes of deeper queues on the ask (resp., bid) side enter the computation of the queue imbalance, 
and this will likely reset the signal.  
As noted in \cite{MP18sta}, this interaction can be deemed responsible for the mean-reverting behaviour of price dynamics in high-frequency markets. 

Moreover, 
we use $\stateVariable_2$ to reproduce 
the update of the limit order book configuration
that happens when a labelled agent submits their market orders.
Indeed, 
we consider normalised volumes up to level $n$, 
namely we assume that $\sum_{i=1}^{n} (\nthBestBidVolume[i]_t + \nthBestAskVolume[i]_t) \equiv 1$,
and we assume that the 
$2n$-tuple $(\nthBestAskVolume[1]_t,\nthBestBidVolume[1]_t, \dots, \nthBestAskVolume[n]_t,\nthBestBidVolume[n]_t)$ 
is distributed as a Dirichlet random variable with $2n$-dimensional parameter
$\dirichletParam=\dirichletParam(\stateVariable(t)) \in \R_{+}^{2n}$ 
that depends on the state variable at time $t$. 

Given the time evolution of the limit order book in the time window $\timeWindow$, 
an estimator for $\dirparam(x)$, 
with $x$ ranging from $1$ to $3K$, 
can be obtained by maximum likelihood estimation. 
Once $\dirparam$ is known, the order book mechanics can be reproduced 
by drawing from the conditional distribution 
$\text{Dir}_{\dirichletParam(\stateVariable(t))}(\cdot \vert \stateVariable_2 (t) )$.
This is the Dirichlet distribution of  the $2n$-tuple 
$(\nthBestAskVolume[1]_t,\nthBestBidVolume[1]_t, \dots, \nthBestAskVolume[n]_t,\nthBestBidVolume[n]_t)$ with parameter $\dirparam(\stateVariable(t))$ conditioned on 
\begin{equation*}
 \frac{2\stateVariable_2(t) - K - 1}{2K}
 \leq 
 \underbrace{\sum_{i=1}^{n} \left( \nthBestBidVolume[i]_t - \nthBestAskVolume[i]_t \right)}_{= \volumeImbalance^n_t}
 <
 \frac{2\stateVariable_2(t) +1}{K}.
\end{equation*}

Algorithm \ref{algo.lob_update} describes how to reproduce 
the order book update in the case of the arrival of a sell market order $(t,q_M, 0, -1)$.
The case of buy market orders is analogous.

\begin{algorithm}[h]
 \caption{State update via order book mechanics (sell market order)}
 \label{algo.lob_update}
 \begin{algorithmic}[5]
  \REQUIRE $\stateVariable(t-)$, $(t,q_M, 0, -1)$
  \STATE set $\ell := (2\stateVariable_2(t-) - K - 1)/(2K)$
  \STATE set $u:= (2\stateVariable_2(t-) +1)/K$
  \STATE sample $V_{t-} = (\nthBestAskVolume[1]_{t-},\nthBestBidVolume[1]_{t-},\dots,\nthBestAskVolume[n]_{t-},\nthBestBidVolume[n]_{t-}) \sim \text{Dir}_{\dirichletParam(\stateVariable(t-))}(\cdot \vert \stateVariable_2 (t-) )$ 
  \STATE set $\stateVariable_1(t) := -1$ if $q_M \geq \nthBestBidPrice[1]_{t-}$; $0$ otherwise
  \STATE initialise $q:=0$
  \FOR {$i$ in $1,\dots,n$}
  \STATE set $v:= \min_{+} (\nthBestBidVolume[i]_{t-}, q_M - q )$
  \STATE set $\nthBestBidVolume[i]_{t} := \nthBestBidVolume[i]_{t-} - v$
  \STATE update $q\leftarrow q+v$
  \ENDFOR
  \STATE set $B:= \sum_{i=1}^{n} \nthBestBidVolume[i]_t$
  \STATE set $A:= \sum_{i=1}^{n} \nthBestAskVolume[i]_{t-}$
  \STATE set $I:= (B-A)/(B+A)$ 
  \STATE set $\stateVariable_2(t):= (2k-K+1)/2$ if $(k-K)/K \leq I < (2k+2 -K)/K$, where $k=0,\dots,K-1$
  \RETURN $(\stateVariable_1(t),\stateVariable_2(t))$
 \end{algorithmic}
\end{algorithm}

Line 4  in Algorithm \ref{algo.lob_update} 
says that the bid price (and consequently the mid-price) decreases 
if the size of the sell market order is larger than the available liquidity 
sitting on the first bid level.
Lines 6:10 cancel from the bid queues the orders whose execution has been triggered by the arrival of $(t,q_M,0,-1)$.
On line 7 we used the notation $\min_{+}(a,b) = \max(0,\min(a,b))$ for $a$ and $b$ real numbers.

\section{Price impact profiles}\label{sec.price_impact_profiles}
Measuring price impact requires two things.
The first is to modify the  model $(\multiCountingProc,\stateVariable)$ of Section \ref{sec.sdhawkes_model} in a way to account for a labelled agent,
whose impact we wish to measure.
The second is to extrapolate to which extent the labelled agent is responsible for the evolution of the price dynamics that emerge from the state process $(\stateVariable(t))_{t}$.
Section \ref{sec.labelled_agent} describes the former; 
Section \ref{sec.impact_a_la_BM} describes  the latter.

\subsection{Labelled agent}\label{sec.labelled_agent}
We account for a labelled agent in the market,
and we aim to measure their impact on the dynamics of the order book.
We take the perspective of a liquidation,
namely we consider our agent (also referred to as liquidator) to be selling the amount $\metaorderSize$ of asset.
The case of acquisition is mutatis mutandis the same.

We let $\timeWindow$
represent the time window of the liquidation.
The quantity $\metaorderSize$
is referred to as the size of the liquidator's metaorder,
or their initial inventory,
and we normalise it
with respect to the overall volume 
$\sum_{i=1}^{n}(\nthBestAskVolume[i]_{0} + \nthBestBidVolume[i]_{0})$
of offers sitting on the first $n$
levels of the order book  at the start of the liquidation window.

We assume that the liquidator intervenes in the market only by sending sell market orders;
they will never place a limit order to queue on the ask side,
but they will initiate trades with existing offers on the bid side.

Hence,
the liquidation is described by the sequence 
$\lbrace (\arrivalTimes[0]_j,
q_{M,j},0,-1): \,
\,
j=1,2,\dots \rbrace$ 
of sell market orders sent by the liquidator.
For every $j$,
$\arrivalTimes[0]_j$ is the time stamp of the liquidator's $j$-th child market order,
and $q_{M,j}$ is its size.

We suppose that the stream of random times 
$\arrivalTimes[0]_1 < \arrivalTimes[0]_2 < \dots$ is confined in $\timeWindow$.
We assume non-explosiveness,
so that the number of liquidator's market orders is finite if the time horizon $\timeHorizon$ of the execution window is not $+\infty$.
Moreover,
we let $\initialTime = \arrivalTimes[0]_1$ represent the time at which the liquidator begins their intervention in the market,
and we let
$\terminationTime:=\sup\lbrace\arrivalTimes[0]_j\leq\timeHorizon: \,
j=1,2,\dots\rbrace$
be the time at which the liquidation stops.

\begin{assumption}\label{assumption.termination-time_of_liquidation}
Let $\metaorderSize$ be the size of the liquidator's metaorder, and for $j$ in $\N$,  let $z_j = q_{M,1}+\dots+q_{M,j}$ be the sum of all liquidity-normalised sizes of the first $j$ child market orders sent by the liquidator. Then, the termination time $\terminationTime$ of the liquidation is assumed to coincide with the smallest time stamp $\arrivalTimes[0]_j$ among the liquidator's market orders such that $z_j \geq \metaorderSize$, namely
 \begin{equation*}
\terminationTime = \inf\left\lbrace \arrivalTimes[0]_j \geq \initialTime : \, \, 
 \sum_{k=1}^{j} q_{M,k} \geq \metaorderSize \right\rbrace.
 \end{equation*}
\end{assumption}

We introduce the liquidator's presence in the model described in Section \ref{sec.sdhawkes_model} by expanding the dimension of the counting process $\multiCountingProc$:
we let the zero-th component $\countingProc[0](t)$ count the liquidator's market orders sent to the exchange by time $t$.
In other words, from the overall sequence $(\arrivalTimes[1]_j)_j$ of arrival times of market orders described in Section \ref{sec.sdhawkes_model},
we extract those sent by the liquidator and we label them as $(\arrivalTimes[0]_j)_j$;
we then let 
\begin{equation*}
 \countingProc[0](t):=\sum_{j\geq 1}\one \left\lbrace \arrivalTimes[0]_j \leq t \right\rbrace
\end{equation*}
count the number of trades initiated by the liquidator that happened before or at time $t$.
Notice that the map $t\mapsto \countingProc[0](t)$ represents how the liquidator is splitting in time the execution of their metaorder.
In other words, this is 
the liquidator's execution schedule.

The pair $(\multiCountingProc, \stateVariable)$ is a state-dependent Hawkes process where the counting process component $\multiCountingProc$ is five-dimensional, 
and the state process $\stateVariable$ is as in equation \eqref{eq.state-variable}.
The event types will be labelled $e=0,1,\dots, 4$ 
and the states will be labelled $x=1,\dots,3K$ or $x=(x_1,x_2)$ with $x_1 = -1,0,+1$ and $x_2 = -(K-1)/2,\dots,+(K+1)/2$.
 The following assumption is in place on the intensities.
\begin{assumption}\label{assumption.imp_and_dec_coef_from_liquidator}
 For all $e=1,\dots,4$, the Hawkes kernel $\hawkesKernel_{0,e}$ coincides with  $\hawkesKernel_{1,e}$.
\end{assumption}

Assumption \ref{assumption.imp_and_dec_coef_from_liquidator}
guarantees consistency in the effect that trades have on the order book dynamics.
It says that the rates of arrival of market orders to the exchange are modified by the liquidator's interventions in the same way as they are by other participants' sell market orders.
More precisely, for $e=1,\dots, 4$ it holds 
\begin{equation}\label{eq.intensities_when_liquidator_is_present}
\begin{split}
\intensity[e](t) =& \baseRate_e 
 + \sum_{\eone = 0}^{4}\int_{[0,t)} \hawkesKernel_{\eone,e}(t-s,\stateVariable(s))d\countingProc[\eone](s)
 \\
 =& \baseRate_e 
 + \sum_{\eone = 1}^{4}\int_{[0,t)} \hawkesKernel_{\eone,e}(t-s,\stateVariable(s))d\countingProc[\eone](s)
 + \int_{[0,t)} \hawkesKernel_{1,e}(t-s,\stateVariable(s))d\countingProc[0](s).
 \end{split}
\end{equation}

The liquidator's execution schedule admits an absolutely continuous compensator $\compensator_{0}$ with density 
\begin{equation}\label{eq.intensity_of_liquidator}
 \intensity[0](t)
 =
 \baseRate_0 \one_{[0,\terminationTime)}(t) + 
 \sum_{\eone,\xone}\one_{[0,\terminationTime)}(t)
 \int_{[0,t)} \hawkesKernel_{\eone,0} (t-s,\xone) d\hybridHawkes_{\eone,\xone}(s).
\end{equation}


For $j=1,2,\dots$ let $(\arrivalTimes[0]_j,
q_{M,j},0,-1)$ be the liquidator's child market orders,
as denoted above.
The liquidator's order  scheduling depends on 
the Hawkes parameters $\baseRate_0$,
and $\hawkesKernel_{\eone,0}$,
which modulate the sequence of arrival times $\arrivalTimes[0]_j$.
Additionally,
the liquidation depends on
the size $q_{M,j}$ of the $j$-th child market order,
for $j = 1, 2, \dots$.\footnote{
Every $q_{M,j}$ satisfies 
the measurability constraint $q_{M,j} \,
\hat{\in} \,
$ $ \filtrationF_{\arrivalTimes[0]_j -} = \bigvee_{\epsilon>0}\filtrationF_{\arrivalTimes[0]_j -\epsilon}$,
where $\filtrationF$ is a history of $(\multiCountingProc,\stateVariable)$.
}
The evolution of the limit order book is simulated by combining
Algorithm \ref{algo.MP18_ogata} and
Algorithm \ref{algo.lob_update},
as detailed in Algorithm \ref{algo.lob_with_liquidator}.

\begin{algorithm}[h]
\caption{Simulation of order book in the presence of liquidator}
\label{algo.lob_with_liquidator}
 \begin{algorithmic}[5]
  \REQUIRE $(\arrivalTimes[]_i,\event[i],\stateVariable_i)_{i=1,\dots,n-1}$, $q_M$
  \STATE \textbf{do} Lines 1:8 of Algorithm \ref{algo.MP18_ogata}
  \STATE draw $\event[n]$ with probabilities proportional to 
	 $\lbrace \intensity[0](\arrivalTimes[]_n),\intensity[1](\arrivalTimes[]_n),\dots, \intensity[4](\arrivalTimes[]_n)\rbrace$
  \IF {$\event[n] = 0$}
  \STATE set $\stateVariable_n \leftarrow$ Algorithm \ref{algo.lob_update}($\stateVariable_{n-1}$,$(\arrivalTimes[]_n,q_M,0,-1)$)
  \ELSE
  \STATE draw $\stateVariable_n$ in $\lbrace 1,\dots,3K\rbrace$ with probabilities $\lbrace \transProb_{\event}(\stateVariable_{n-1},1), \dots,\transProb_{\event}(\stateVariable_{n-1},3K)\rbrace$
  \ENDIF
  \RETURN $(\arrivalTimes[]_n,\event[n],\stateVariable_n)$
 \end{algorithmic}
\end{algorithm}

\begin{remark}\label{remark.BM_execution_schedule}
In the aforementioned \cite{BM14haw} (see Remark \ref{remark.BM14_model_comparison}),
a labeled agent is accounted for by considering 
the following intensities of the four-dimensional counting process $\countingProc[]$.
For $e=1,\dots,4$ and $t>0$ they set 
\begin{equation}\label{eq.BM14_intensity_with_liquidator}
 \intensity[e](t) = 
 \baseRate_e
 +\sum_{\eone=1}^{4}\int_{[0,t)} \hawkesKernel\subscriptee (t-s)d\countingProc[\eone](s) 
 + \int_{[0,t)} \theta_e(s)dA(s),
\end{equation}
where 
$t\mapsto A(t)$ is the liquidator's execution schedule,
$t\mapsto \theta_1(t)$ 
(resp., $t\mapsto\theta_2(t)$)
represents the impact of the liquidator's market orders on the arrival of other participants' sell (resp., buy) market orders,
and $t\mapsto \theta_3(t)$ 
(resp., $t\mapsto \theta_4(t)$) 
represents the impact of the liquidator's market orders on downward (resp., upward) jumps of the mid-price.
In their model, to have consistency between the liquidator and other market participants one needs to impose
$\theta_3(t) = \hawkesKernel_{1,3}(t)$ 
and 
$\theta_4(t)=\hawkesKernel_{1,4}(t)$ for $t\geq 0$.
Practically, 
this implies that the atomic components in the Hawkes kernel 
are passed to the integrands $\theta_3$ and $\theta_4$,
which means that 
the liquidator walks the book at an average rate equal to the overall proportion of 
markets orders walking the book. The consequence that the liquidator walks the book in this way can be a potentially undesirable feature.\footnote{To see why this can be a potentially undesirable feature, consider a liquidator that trades with child orders whose sizes are significantly different from the average market orders arriving in the market.}  In our model, there is no need for this to be assumed. We are able to test executions without this assumption, 
and still maintain consistency between the liquidator and other market participants.
\end{remark}

\begin{remark}\label{remark.why_liquidator_is_hawkes}
The liquidator's interventions in the market
have been modelled 
by expanding one component of the Hawkes process 
introduced in Section \ref{sec.sdhawkes_model}.
The justification for this modelling choice is twofold. 
First, this guarantees consistency between 
executions of sell market orders sent by the liquidator
and executions of sell market orders sent by other market participants.
Given our interest in understanding the liquidator's impact,
any other stochastic model would raise questions of granting the liquidator 
with a privileged order scheduling.
Second, expanding the dimensions of $\countingProc[]$ allows us to give 
a natural justification to \cite{BM14haw}'s formula
for the intensity \eqref{eq.BM14_intensity_with_liquidator} -- 
$\hawkesKernel_{0, e}$ takes the role of $\theta_e$ and $dN_0$ takes the role of $dA$.
Hence, our modelling choice resonates with existing models in the literature, and it is grounded in our phenomenological point of view. A future work could adopt the point of view of optimal execution, and optimise  the liquidator's scheduling in a set of admissible liquidation strategies aimed at minimising their price impact. 
\end{remark}

\subsection{Definition of price impact}
\label{sec.impact_a_la_BM}
We partition the state space $\stateSpace$ according to
the values of the first component $\stateVariable_1$ of the state variable $\stateVariable= (\stateVariable_1, \stateVariable_2)$.
We define
\begin{equation}\label{eq.partition_of_state_space}
\begin{split}
 \stateSpace^{x_1} := & \left\lbrace y=(y_1,y_2) \in \stateSpace:
\, \, y_1 = x_1\right\rbrace.
\end{split}
\end{equation}
We refer to states $x$ in $\stateSpace^{+}$ 
(resp., in $\stateSpace^{-}$) 
as inflationary (resp., deflationary) states.

The jump times for the mid-price consequently give rise to the counting processes 
\begin{equation}\label{eq.jump_times}
 \countingProc[]^{x_{1}}(t) := \sum_{n}\one\left\lbrace \arrivalTimes[]_n\leq t , \stateVariable_1(\arrivalTimes[]_n) = x_1 \right\rbrace,
 \qquad x_1 \in \lbrace-1,0,+1\rbrace,
\end{equation}
where $\arrivalTimes[]_n$ is the $n$-th jump time of the ground process.
The difference 
$
 \countingProc[]^{+}(t) - \countingProc[]^{-}(t)
$
is a proxy for the mid-price in the order book.
Indeed, 
we can rewrite the integral quantity in equation \eqref{eq.midprice_proxy} as 
\begin{equation*}
 \intzerot \stateVariable_1 (s)d\groundProc(s)
 =
 \countingProc[]^{+}(t) - \countingProc[]^{-}(t).
\end{equation*}

\begin{defi}\label{def.price-simmetry}
 The state-dependent Hawkes model $\sdHawkesPair$ is said price-symmetric if 
 for all $t\geq0$
	\begin{equation*}
		\left(
		\sum_{x\in\inflationarySpace} - \sum_{x\in\deflationarySpace}
		\right)
		\sum_{e=1}^{4} \transProb_{e}\left(\stateVariable(t), x\right)
		\ell_{e}(t)=0,
	\end{equation*}
 where 
\begin{equation*}
 \ell_{e}(t) = \baseRate_{e}
 +
 \sum_{\eone=1}^{4}\int_{[0,t)} \hawkesKernel\subscriptee(t-s,\stateVariable(s)d\countingProc[\eone](s).
\end{equation*}
\end{defi}

\begin{prop}\label{prop.sufficient_cond_price_symmetry}
	Assume that there exist 
	a permutation $\sigma_E$ of $\lbrace1,\dots, 4\rbrace$
	and a bijective map $\sigma_S: \inflationarySpace\rightarrow\deflationarySpace$
	such that
 \begin{enumerate}[label={(\emph{\roman{*}})} , ref={\ref{prop.sufficient_cond_price_symmetry} (\emph{\roman{*}})}]
 \item\label{prop.price-simmetry.trans_prob}
	 $\transProb_{e}(y,x) = \transProb_{\sigma_E(e)}(y,\sigma_S(x))$
		for all $y$ in $\stateSpace$, all $x$ in $\inflationarySpace$ 
		 and all $e=1,\dots,4$;
 \item\label{prop.price-simmetry.base_rates}
	 $\baseRate_{e} = \baseRate_{\sigma_E(e)}$
		 for all $e=1,\dots,4$;
 \item\label{prop.price-simmetry.kernels}
	 $\hawkesKernel\subscriptee(t, \xone) = \hawkesKernel_{\eone, \sigma_E(e)}(t, \xone)$
		 for all $\xone$ in $\stateSpace$ all $e, \eone = 1, \dots, 4$ 
		 and all $t\geq0$.
\end{enumerate}
Then, $\sdHawkesPair$ is price-symmetric.
\end{prop}

\begin{remark}
The condition in Proposition \ref{prop.price-simmetry.trans_prob}
captures the idea that, given the current state $y$, 
transitions to inflationary states  and transitions to deflationary states  are equally likely.
The conditions in Proposition \ref{prop.price-simmetry.kernels}
capture the idea that, given the current state
$y$, every event-state pair
$(\eone,\xone)$
excites an  event-state pair $(e,x)$
with inflationary state $x\in\inflationarySpace$
 the same way as 
it excites an event-state pair $(\sigma_{E}(e),\sigma_{\stateSpace}(x))$
with deflationary state $\sigma_{\stateSpace}(x) \in\deflationarySpace$; 
in other words, the offspring from every event-state pair $(\eone,\xone)$
are equally likely to be associated with inflationary states or with deflationary states. 
\end{remark}

\begin{defi}\label{def.BM_impact}
Let $t_0$ be the time when the liquidator becomes active in the market.
Then,
the price impact profile of the execution schedule $\countingProc[0]$
is the primitive of
$t\mapsto \directImpact(t) + \indirectImpact(t)$
pinned at $0$ in $t_0$, 
where
\begin{equation*}
\directImpact(t)=  
\sum_{x\in\deflationarySpace}\transProb_{0} (\stateVariable(t),x)
\left(\baseRate_{0}  
+ \sum_{\eone=1}^{4}\sum_{\xone=1}^{3K} \int_{[0,t)} \hawkesKernel_{\eone,0}(t-s,\xone)d\hybridHawkes_{0,\xone}
\right)
\one_{[0,\terminationTime)} (t),
\end{equation*}
where $\terminationTime$ is the termination time of the liquidation,
\begin{equation}\label{eq.transProb_liquidator}
\transProb_{0}(\xone,x) = 
\frac{\sum_j \one\lbrace\stateVariable(\arrivalTimes[0]_j - )=\xone, \, \, \stateVariable(\arrivalTimes[0]_j )=x \rbrace} {\sum_j \one\lbrace\stateVariable(\arrivalTimes[0]_j - )=\xone \rbrace},
\end{equation}
and
\begin{equation*}
\indirectImpact(t)=
\sum_{e=1}^{4}\sum_{\xone=1}^{3K} \int_{[0,t)} \hawkesKernel_{1,e}(t-s,\xone)d\hybridHawkes_{0,\xone}
\left(
\sum_{x\in\deflationarySpace}
-
\sum_{x\in\inflationarySpace}
\right) \transProb_{e}(\stateVariable(t),x).
\end{equation*}
The map
$t\mapsto \directImpact(t) + \indirectImpact(t)$
is referred to as intensity of the price impact profile.
\end{defi}

\begin{remark}
The intensity of the price impact profile is decomposed in two components,
namely $\directImpact(t)$
and $\indirectImpact(t)$.
Both are null if $\countingProc[0] (t) \equiv 0$.
The former is referred to as ``direct'' impact
and stems from those summands of the execution schedule's intensity $\intensity[0](t) = \sum_{x=1}^{3K}\hybridIntensity_{0,x}(t)$ that are associated with deflationary states,
namely $\directImpact(t) = \sum_{x\in\deflationarySpace} \hybridIntensity_{0,x}(t)$.
Notice that $\directImpact(t)\geq 0$ for all $t>0$
and  $\directImpact(t) = 0$ for all $t>\terminationTime$.
On the contrary,
the second term $\indirectImpact(t)$ stems from events originated by participants other than the liquidator but in response to the liquidator's interventions,
hence the name of ``indirect'' impact.
It can have either sign
and it is in general non-zero even beyond the termination time;
for this reason it is linked to the transient impact.
More precisely,
for $t>\terminationTime$ it holds 
$
\indirectImpact(t) 
= 
\sum_{e=1}^{4}\sum_{\xone} 
\sum_{j} \hawkesKernel_{1,e}(t-\arrivalTimes[0]_j,\stateVariable(\arrivalTimes[0]_j)) 
(\sum_{x\in \deflationarySpace} - \sum_{x\in \inflationarySpace})
\transProb_{e}(\stateVariable(t),x)$,
and  the transient price impact profile is the map 
$t\mapsto \intzerot \indirectImpact(s)ds$,
restricted to the interval
$t\geq \terminationTime$.
\end{remark}

In a price-symmetric state-dependent Hawkes model,
if $\countingProc[0]\equiv0$, then 
$\Nminus - \Nplus$ is a martingale, 
and its compensator is identically null. 
Instead, 
when the liquidator is active in the market,
the symmetry is disrupted, 
and we map this disruption to our measure of the price impact.

Hence, 
Definition \ref{def.BM_impact} is vindicated by the following proposition.

\begin{prop}\label{prop.bm_impact}
	If $\sdHawkesPair$ is price-symmetric,
	then
	the price impact profile of $\countingProc[0]$ is the 
	$\internalHistory$-compensator of $\countingProc[]^{-} - \countingProc[]^{+}$,
	where $\internalHistory$ is the minimal complete right-continuous filtration to which $\sdHawkesPair$ is adapted. 
\end{prop}

The direct impact component $\directImpact(t)$ 
of the intensity $(\intensityDeflationary - \intensityInflationary)(t)$ 
encompasses the transition matrix $\transProb_0$ 
associated with the state update that occurs when liquidator's orders are executed.
For $x\derivative$
and $x$ in $\stateSpace$,
$\transProb_{0}(x\derivative,x)$ is estimated according to equation \eqref{eq.transProb_liquidator};
hence it summarises the state transitions that stem from Algorithm \ref{algo.lob_update} during the simulation of the execution.
This disentangles the effects of liquidator's orders (whose sizes $q_{M,j}$ are set by the liquidator) from the effects of other market orders,
i.e.
$\transProb_0 \neq \transProb_1$ in general,
allowing to investigate the impact of different execution strategies.

In particular,
the liquidator might choose to send market orders with sizes that
never exceed the available liquidity on the first bid level;
this would cause $\transProb_0(x\derivative,x)=0$ for all $x\derivative$ in $\stateSpace$
and all deflationary $x$ in $\deflationarySpace$,
and thus $\directImpact(t)\equiv 0$.
Nonetheless,
the overall impact would not be null,
because of the indirect term $\indirectImpact(t)$.
Indeed,
even without ever walking the book,
the liquidator's orders would modify
(\emph{i})
the arrival of orders submitted by other market participants who react to 
the liquidator's executions;
(\emph{ii})
the volumes in the order book.

As far as (\emph{i}) is concerned,
if the dynamics of order submission is such that
deflationary events trigger other events with deflationary effects on the price,
then the price may plunge as an indirect consequence of the liquidation.

As far as (\emph{ii}) is concerned,
despite the fact that they do not walk the book,
liquidator's executions consume liquidity on the bid side, 
pushing the state trajectory $t\mapsto \stateVariable(t)$ to dwell in states 
$\lbrace y=(y_1,y_2)\in\stateSpace:
\,
\,
y_2<0\rbrace$
for longer.
The probability of transitioning from these states to deflationary states is higher than the probability of transitioning to inflationary states,
hence making the term
$(\sum_{x\in\deflationarySpace} - \sum_{x\in\inflationarySpace})\transProb_{e}(\stateVariable(t),
x)$
positive,
and contributing to the impact via the indirect term $\indirectImpact(t)$.
Notice that this form of impact would not be captured by a less granular model 
where the update of the volumes in the limit order book is not reproduced as we do in Algorithm \ref{algo.lob_update},
and where the liquidator's child orders 
are assumed to walk the book at an average rate 
equal to the overall proportion of market orders
that do so 
-- see Remark \ref{remark.BM_execution_schedule}.

In Section \ref{sec.applications},
we will see that,
after calibrating our model on empirical data from NASDAQ, 
the indirect component of the price impact is actually 
the main driver of price impact during liquidation.

\section{Applications}\label{sec.applications}
\subsection{Description of the dataset and model specifications}
We study order book data provided by \href{https://lobsterdata.com/}{LOBSTER}.\footnote{See 
\url{https://lobsterdata.com/}
} 
LOBSTER is a provider of high-quality limit order book data that is reconstructed from \href{http://nasdaqtrader.com/Trader.aspx?id=ITCH}{NASDAQ's Historical TotalView-ITCH}\footnote{See 
\url{http://nasdaqtrader.com/Trader.aspx?id=ITCH}
} files with detailed event information. 
The reconstruction methodology is described in \cite{HP11lob}.

For every NASDAQ ticker and every active trading day,
LOBSTER provides two files in \codehl{.csv} format:
a `message file' and an `orderbook' file. 
The former is an event-by-event history of messages sent to the exchange that provoked an update in the configuration of the order book. 
The latter is an event-by-event snapshot of the order book,
where the $n$-th row corresponds to the configuration resulting from the $n$-th message reported in the message file. 

Prices are reported in $10^{-4}$USD; 
hence the tick size, imposed by regulation\footnote{See rule 4701(k) at
\url{https://listingcenter.nasdaq.com/rulebook/nasdaq/rules/nasdaq-4000}
} and equal for all shares with price above 1USD, is set to 100. 
Time stamps are reported in seconds after midnight with resolution at the nanosecond scale. 
Events happening in the trading venue are labelled according to Table \ref{tab.LOBSTERlabels}.\footnote{See
\url{https://lobsterdata.com/info/DataStructure.php}.
}
 
\begin{table}[h!]
\centering
 \begin{tabular}{rl}
1: & Submission of a new limit order \\
2: & Cancellation (partial deletion of a limit order)\\
3: & Deletion (total deletion of a limit order)\\
4: & Execution of a visible limit order\\
5: & Execution of a hidden limit order\\
6: & Indicates a cross trade, e.g. auction trade\\
7: & Trading halt indicator \\
 \end{tabular}
\caption{LOBSTER labels of order book events} 
\label{tab.LOBSTERlabels}
\end{table}

Table \ref{tab.LOBSTERmapping} shows how 
we map LOBSTER order book labels 
to the sequences of arrival times described in Section \ref{sec.sdhawkes_model}.

\begin{table}[h!]
\centering
{
\begin{tabular}{ccc}
    \hline
	\textbf{LOBSTER event label} & \textbf{Bid/Ask} &  \textbf{Event type} \\
    \hline
                       1 &     ask &           3 \\
                       1 &     bid &           4 \\
                       2 &     ask &           4 \\
                       2 &     bid &           3 \\
                       3 &     ask &           4 \\
                       3 &     bid &           3 \\
                       4 &     ask &           2 \\
                       4 &     bid &           1 \\
                       5 &     ask &           2 \\
                       5 &     bid &           1 \\
    \hline
\end{tabular}
}
\caption{Mapping of LOBSTER labels to event types}
\label{tab.LOBSTERmapping}
\end{table}

In the analysis that follows, we study order book data for the ticker INTC trading on January 25, 2019. First, we calibrate our state-dependent Hawkes model on the dataset of January 25, 2019;
then, we simulate liquidations of a large number of shares using Algorithm \ref{algo.lob_with_liquidator}; 
and finally we assess the price impact of such simulated liquidations as per Definition \ref{def.BM_impact}. At the end of the section we make remarks about the sensitivity of our results with respect to calibrated parameters, and we provide insights on the implementation for other dates and tickers.

\subsection{Calibration}\label{sec.calibration}
After filtering for the arrival times $(\arrivalTimes[e]_j)_j$,
$e=1, \dots, 4$, 
and defining the state variable $\stateVariable = (\stateVariable_1, \stateVariable_2)$ as per equation \eqref{eq.state-variable} with $n=2$ and $K=3$,
the data sets of message file and order book 
for INTC on January 25, 2019 are as in Table \ref{tab.filteredDatasets}.

\begin{table}[h!]
\centering
\begin{tiny}
	\begin{tabular}{l}	
{
\begin{tabular}{lcccrr}\\
    \hline
	{} & \textbf{Bid/Ask} &  \textbf{LOBSTER event label} &  \textbf{Event type} &   \textbf{price} &  \textbf{size} \\
    \textbf{time}         &         &                      &             &         &       \\
    \hline
    \textbf{35400.092452} &     ask &                    4 &           2 &  460700 &   300 \\
    \textbf{35400.092533} &     ask &                    1 &           3 &  460700 &   200 \\
    \textbf{35400.092768} &     ask &                    4 &           2 &  460700 &   100 \\
    \textbf{35400.113748} &     ask &                    4 &           2 &  460700 &   400 \\
    \textbf{35400.113776} &     bid &                    1 &           4 &  460700 &   100 \\
    \textbf{35400.121175} &     bid &                    4 &           1 &  460700 &   253 \\
    \textbf{35400.121258} &     ask &                    1 &           3 &  460700 &  3300 \\
    \textbf{35400.123294} &     ask &                    4 &           2 &  460700 &   100 \\
    \textbf{35400.123334} &     ask &                    4 &           2 &  460700 &     6 \\
    \textbf{35400.125010} &     ask &                    4 &           2 &  460700 &    14 \\
    \hline
\end{tabular}
}
\\
\\
{
\begin{tabular}{lcrrrr}\\
    \hline
    {} &   \textbf{State}  &  \textbf{ask price 1} &  \textbf{ask volume 1} &  \textbf{bid price 1} &  \textbf{bid volume 1} \\
	\textbf{time}         &           &            &             &            &        \\    
    \hline
    \textbf{35400.092452} &    (1, 1) &     460700 &         200 &     460600 &         900 \\
    \textbf{35400.092533} &   (-1, 1) &     460700 &         200 &     460600 &        1578 \\
    \textbf{35400.092768} &    (0, 1) &     460700 &         200 &     460600 &        1553 \\
    \textbf{35400.113748} &    (1, 1) &     460700 &         300 &     460600 &        1029 \\
    \textbf{35400.113776} &   (1, -1) &     460800 &        1600 &     460700 &         100 \\
    \textbf{35400.121175} &  (-1, -1) &     460800 &        1800 &     460700 &         153 \\
    \textbf{35400.121258} &   (-1, 0) &     460700 &        3300 &     460600 &        1029 \\
    \textbf{35400.123294} &    (0, 1) &     460700 &         352 &     460600 &         900 \\
    \textbf{35400.123334} &    (0, 1) &     460700 &         346 &     460600 &         900 \\
    \textbf{35400.125010} &    (0, 1) &     460700 &         332 &     460600 &         900 \\
    \hline
\end{tabular}
}
\end{tabular}
\end{tiny}
\caption{Ten time stamps from the filtered message file and order book file. We show the first level of the limit order book for ease of exposition.  }
\label{tab.filteredDatasets}
\end{table}

Starting from these data sets we perform maximum likelihood extimation of our state-dependent Hawkes model. 

Transition probabilities are straightforwardly estimated from empirical frequencies. 
For every event $e=1, \dots, 4$,
we estimate a $9\times9$-transition matrix $\transProb_{e}$
that describes the law of the state-update in equation \eqref{eq.markov_update_stateVariable}.
In Table \ref{tab.calibratedTransProb},
we show the result of this estimation focusing on events of type either 1 or 2, 
i.e. execution on either the bid or the ask side.

\begin{table}[H]
	\centering
	\begin{scriptsize}
	\begin{tabular}{c}
		Probabilities of mid-price movements $X_1(T)\in \lbrace-1, 0, +1\rbrace$ 
		when event of type  either 1 or 2 occurs at  \\
		time $T$ and 
		queue imbalance $X_2(T-)$ is negative (-1), neutral (0) or positive (+1): \\
		\begin{tabular}{crrrr}\\
			& $X_1$ &        \textbf{-1} &         \textbf{0} &         \textbf{1} \\
			\textbf{event} & $X_2$ &           &           &           \\
		    \hline\\
			\multirow{3}{*}{\textbf{\emph{1}}} 
			& \textbf{-1} &  0.476657 &  0.523343 &  0.000000 \\
			&  \textbf{0} &  0.410729 &  0.589271 &  0.000000 \\
			&  \textbf{1} &  0.346157 &  0.653843 &  0.000000 \\
			\cline{1-5}\\
			\multirow{3}{*}{\textbf{\emph{2}}} 
			& \textbf{-1} &  0.000000 &  0.702554 &  0.297446 \\
			&  \textbf{0} &  0.000000 &  0.617759 &  0.382241 \\
			&  \textbf{1} &  0.000000 &  0.533823 &  0.466177 \\
		    \hline
		 \end{tabular}
		 \\
		 \\
		Full transition probability from $X(T-) = (X_1(T-), X_2(T-))$ to $X(T) = (X_1(T), X_2(T))$ 
		when \\
		an execution happens on the bid side at time $T$.\\
		\begin{tabular}{crrrrrrrrrrr}\\
			&    & $X_1$ & \multicolumn{3}{l}{\textbf{-1}} & \multicolumn{3}{l}{\textbf{0}} & \multicolumn{3}{l}{\textbf{1}} \\
			&    & $X_2$ &        \textbf{-1} &         \textbf{0} &         \textbf{1} &        \textbf{-1} &         \textbf{0} &         \textbf{1} &   \textbf{-1} &    \textbf{0} &    \textbf{1} \\
			\textbf{event} & $X_1$ & $X_2$ &           &           &           &           &           &           &      &      &      \\
        \hline\\
			\multirow{9}{*}{\textbf{\emph{1}}} & \multirow{3}{*}{\textbf{-1}} & \textbf{-1} &  0.176622 &  0.175752 &  0.000000 &  0.359614 &  0.283382 &  0.004630 &  0.0 &  0.0 &  0.0 \\
			&    &  \textbf{0} &  0.059468 &  0.279633 &  0.005352 &  0.077036 &  0.550193 &  0.028318 &  0.0 &  0.0 &  0.0 \\
			&    &  \textbf{1} &  0.009081 &  0.217948 &  0.071566 &  0.019519 &  0.433721 &  0.248165 &  0.0 &  0.0 &  0.0 \\
        \cline{2-12}\\
			& \multirow{3}{*}{\textbf{0}} & \textbf{-1} &  0.287448 &  0.245331 &  0.003610 &  0.414895 &  0.048716 &  0.000000 &  0.0 &  0.0 &  0.0 \\
			&    &  \textbf{0} &  0.065844 &  0.350749 &  0.021704 &  0.083769 &  0.473489 &  0.004445 &  0.0 &  0.0 &  0.0 \\
			&    &  \textbf{1} &  0.005626 &  0.162936 &  0.130160 &  0.008709 &  0.242360 &  0.450209 &  0.0 &  0.0 &  0.0 \\
        \cline{2-12}\\
			& \multirow{3}{*}{\textbf{1}} & \textbf{-1} &  0.257898 &  0.279307 &  0.004003 &  0.327795 &  0.130997 &  0.000000 &  0.0 &  0.0 &  0.0 \\
			&    &  \textbf{0} &  0.066807 &  0.357431 &  0.025199 &  0.091674 &  0.447274 &  0.011614 &  0.0 &  0.0 &  0.0 \\
			&    &  \textbf{1} &  0.013147 &  0.224193 &  0.203817 &  0.033174 &  0.353798 &  0.171872 &  0.0 &  0.0 &  0.0 \\
        \hline\\
        \end{tabular}
	\\
	\\	
	\\
	\\
	\\
	\\	
	\end{tabular}
	\end{scriptsize}
	\caption{Transition probabilities $\transProb_{e}$ calibrated on INTC as of January 25, 2019.}
	\label{tab.calibratedTransProb}
\end{table}

\begin{assumption}\label{assumption.powerlaw_kernel}
Hawkes kernels are assumed in the parametric form 
\begin{equation}\label{eq.powerlaw_kernels}
 \hawkesKernel\subscriptee(t,\xone) 
 = \impCoef\subscriptexe \big( t+ 1 \big)^{-\decCoef\subscriptexe},
\end{equation}
for some non-negative coefficients $\impCoef\subscriptexe \geq 0$ and $\decCoef\subscriptexe > 1$. 
\end{assumption}

\begin{remark}\label{remark.why_powerlaw_kernel}
Assumption \ref{assumption.powerlaw_kernel} is grounded in 
the stylized fact that power-law kernels better fit real world data 
than exponential kernels do,
albeit being more computational expensive.
This assumption also builds on \cite{BM14haw}'s findings.
In the aforementioned paper, 
the authors devise a non-parametric estimation for Hawkes kernels.
Once this non-parametric estimation has converged,
they compare the estimated kernels with parametric ones, 
and confirm that indeed the decay of the kernels is of power-law type.
\end{remark}

We estimate the parameters $\baseRate_e$, $\impCoef\subscriptexe$, and $\decCoef\subscriptexe$
using a gradient-descent algorithm. 
In Table \ref{tab.estimatedHawkesParam},
we present the result of this estimation 
by reporting the four dimensional vector $\baseRate$,
and $\Lone$ norms of the kernels $\hawkesKernel\subscriptee$.

\begin{table}[H]
	\centering
	\begin{scriptsize}
	\begin{tabular}{c}
		Base rates $\baseRate_e$, $e=1,\dots, 4$. 
		\\
		\begin{tabular}{lrrrr}\\
			{} &   \textbf{event \emph{1}} &   \textbf{event \emph{2}} &   \textbf{event \emph{3}} &   \textbf{event \emph{4}} \\
		    \hline\\
			\textbf{base rate} &  0.040201 &  0.050182 &  0.000735 &  0.000608 \\
		    \hline
                \end{tabular}
		\\
		\\
		\\
		\\
		$L^1$ norms
		$
		\Vert \kappa_{e^{\prime}, e} (\cdot, x) \Vert,
		$
		for
		$
		e = 1, \dots, 4
		$
		corresponding to executions
		on the bid \\ side (event $e^{\prime} = 1$)
		and on the ask side (event $e^{\prime} = 2$),
		bucketed by queue imbalance $X_2 \in \lbrace -1, 0, 1\rbrace$
		\\
		\begin{tabular}{crrrrr}\\
                  & \textbf{Price} \textbf{pressure} &      \textbf{defl} &      \textbf{infl} &      \textbf{defl} &      \textbf{infl} \\
                  & \textbf{event} $e$ &         \textbf{\emph{1}} &         \textbf{\emph{2}} &         \textbf{\emph{3}} &         \textbf{\emph{4}} \\
                \textbf{event} $e\derivative$ & $X_2$ &           &           &           &           \\
                \hline\\
			\textbf{\emph{1}} & \textbf{\emph{-1}} &  1.845517 &  0.332669 &  1.458517 &  0.736039 \\
			&  \textbf{\emph{0}} &  1.851492 &  0.391112 &  1.465794 &  0.751008 \\
			&  \textbf{\emph{1}} &  1.858869 &  0.450124 &  1.457819 &  0.741827 \\
			\textbf{2} & \textbf{\emph{-1}} &  0.311372 &  1.884085 &  0.580953 &  1.358946 \\
			&  \textbf{\emph{0}} &  0.311271 &  1.871763 &  0.604899 &  1.368861 \\
			&  \textbf{\emph{1}} &  0.243171 &  1.874398 &  0.594630 &  1.365085 \\
                \hline
                \end{tabular}
	\end{tabular}
	\end{scriptsize}
	\caption{Hawkes parameters $\baseRate_e$, $\impCoef\subscriptexe$, and $\decCoef\subscriptexe$ calibrated on INTC as of January 25, 2019.}
	\label{tab.estimatedHawkesParam}
\end{table}

Finally, Table \ref{tab.goodness_of_fit} shows QQ plots for goodness-of-fit diagnostics; we note that the fit is adequate for our purposes although there is some deviation in the tail part of the plots -- perfecting the fit tends to be difficult with the amount of data we employ (e.g., for INTC on January 25, 2019 we employ $1,563,582$ datapoints).
Moreover, on two different time-scales,
we visually compare the trajectory of the mid-price 
as reported in LOBSTER, 
as reconstructed from equation \eqref{eq.midprice_proxy},
and as simulated in the calibrated model (one sample).

\begin{table}[H]
	\begin{tabular}{c}
		We test that the time-changed inter-arrival times of equation \eqref{eq.time-changed_intertimes}
		are i.i.d. samples \\
		from a unit rate exponential distribution.
		\\
		\includegraphics[width=0.9\textwidth]{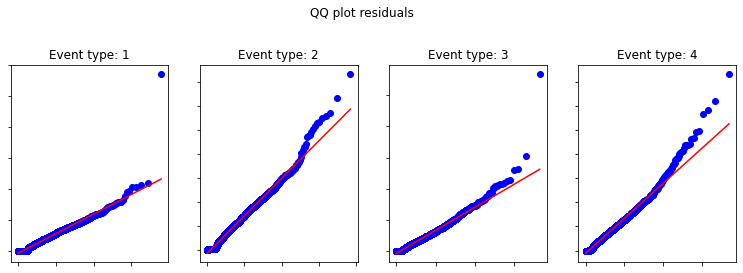}
		\\
 		\\
		Mid-price trajectories on two different time scales. 
		Origin of time is set at 9.55am\\
		on January 25, 2019. Time is measured in seconds.
		\\
		\includegraphics[width=0.45\textwidth]{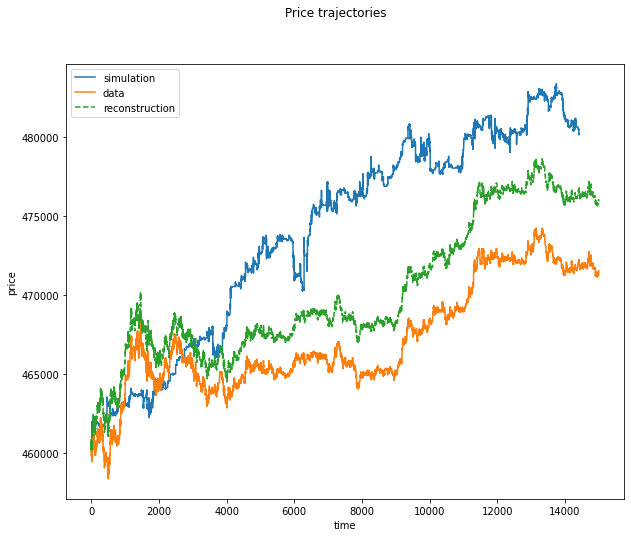}
		\includegraphics[width=0.45\textwidth]{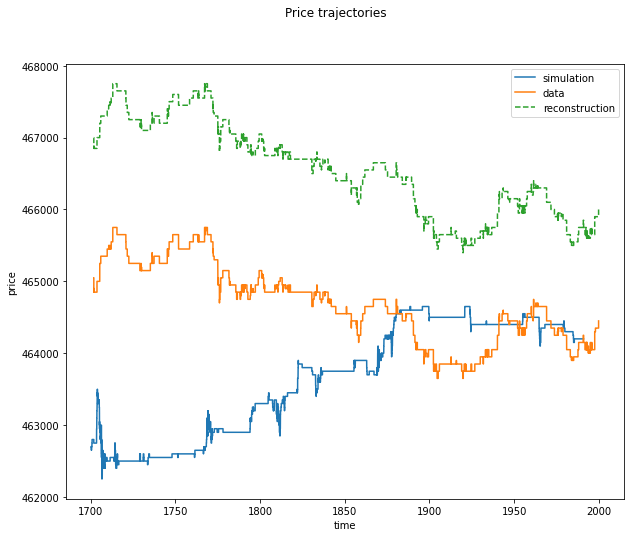}
	\end{tabular}	
	\caption{Goodness-of-fit diagnostics for the model calibrated on INTC as of January 25, 2019.}
	\label{tab.goodness_of_fit}
\end{table}


\subsection{Price impact assessment}

We simulate liquidations 
in our state-dependent Hawkes model for an order book 
calibrated on
LOBSTER data for the ticker INTC on January 25, 2019,
and we assess the price impact of such liquidations
using Definition \ref{def.BM_impact}.

We investigate two aspects of the liquidation schedule:
The rate with which liquidator's orders walk the book
(captured by the transition matrix $\transProb_{0}$),
and 
the clustering of the liquidator's orders in response 
to events happenning in the limit order book.
We modulate these two aspects through 
the parameters reported in Table \ref{tab.liquidation_param}.

\begin{table}[H]
\centering
        {
	\begin{tabular}{lll}
	\hline
		\textbf{Name} & \textbf{Symbol} & \textbf{Formula} \\
	\hline
		base rate & $\baseRate_0$ &  
		Zero-th component of $\baseRate = (\baseRate_0, \baseRate_1, \dots, \baseRate_4)$ \\
		clustering rate & $\clusteringRate$ & 
		$\impCoef_{\eone, x, 0} = \clusteringRate \cdot \impCoef_{\eone, x, 1}$\\
		order size & $\ordersizing$ &
		$q_{M,j} = \ordersizing \cdot \sum_{i=1}^{n}\nthBestBidVolume[i]_{\arrivalTimes[0]_j - } $
	\end{tabular}
	}
	\caption{Parameters of the liquidation schedule}
	\label{tab.liquidation_param}
\end{table}

We run simulations for different values of these parameters
and 
we find out that 
the clustering of liquidator's orders
has a bigger price impact than 
the rate with which they walk the book. 
This suggests that
the dynamic evolution 
of the order book plays a bigger role 
in price formation than 
the instantaneous states of the queues.

Tables \ref{tab.low_walking}, \ref{tab.high_walking} and \ref{tab.with_the_market} 
present
three simulations representative of our findings.

More precisely, 
Table \ref{tab.low_walking}
shows a liquidation in which 
there is no clustering of  market orders
(i.e. $\impCoef_{\eone, x, 0} = 0$),
and the execution schedule follows a Poisson process with rate 
$\baseRate_0 = 0.03$, namely approximately 75\% 
of the base rate of all other sell market orders.

All liquidator's orders have size equal to 7.5\% of the volumes
of bid offers queing on the first $n$ levels of the bid side
at the moment of the order submission. 
This entails that liquidator's orders will 
walk the book only when the aggregate size of level 1 is 
less than 7.5\% of the overall bid volume. 
As a result, the estimated transition matrix 
$\transProb_{0} = \transProb_{0}(x\derivative,x)$
concentrates the mass on those states $x=(x_1, x_2)$ such that 
$x_2 = 0$.
The more positive the queue inbalance 
at the moment of the order submission is,
the more this concentration holds. 

The liquidation takes approximately 8300 seconds to complete. 
The price impact score, defined as the maximum of the price impact profile divided by the duration, 
is 4.11\%. 

The line charts in Table \ref{tab.low_walking} (top left plot) present visualisations of
the liquidation schedule 
(the red dots represent executions on the bid side triggered by one of liquidator's sell market orders), 
and its intensity (see equation \eqref{eq.intensity_of_liquidator}), 
which in this case is simply $\intensity[0](t) = \baseRate_0  \one_{[0,\terminationTime)}(t)$. This panel also depicts the impact profile trajectory (green line in top left plot) that we used to compute the price impact score.
Moreover, 
the trajectories of liquidator's inventory and of the mid-price during execution are plotted (see bottom left plot). 
The latter can be compared to the mid-price simulated when the liquidator was not present in the market 
(see Table \ref{tab.goodness_of_fit}),
and it provides a graphical representation of price impact.

\begin{table}[H]
\centering
	\begin{scriptsize}
	\begin{tabular}{c}
		\begin{tabular}{lr}
			Initial inventory $Q_0$: & 10.0 \\
			Base rate $\baseRate_0$: & 0.03 \\
			Clustering rate $\clusteringRate$: & 0.0 \\
			Order size $\ordersizing$: & 0.075 \\
			Start time $t_0$: & 0.0\\
			Termination time $\tau$: & 8315.9\\
			Price impact score: & 0.04113\\
		\end{tabular}
	\\
	\\
		\includegraphics[width=0.45\textwidth]{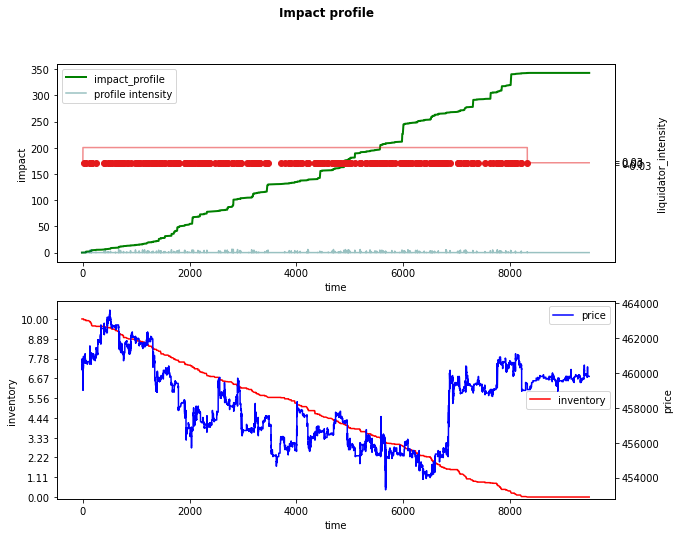}
		\includegraphics[width=0.45\textwidth]{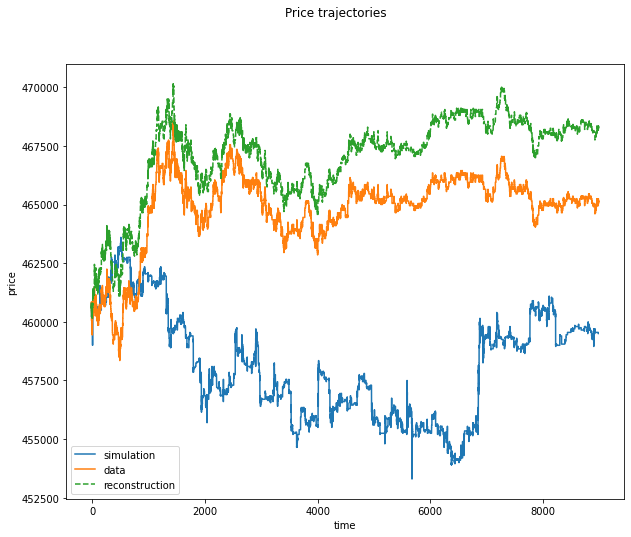}
	\\
	\\	
	\\	
	\\	

		Probabilities of mid-price movements $X_1(T)\in \lbrace-1, 0, +1\rbrace$ 
		when event of type  0, 1 or 2 occurs at  \\
		time $T$ and 
		queue imbalance $X_2(T-)$ is negative (-1), neutral (0) or positive (+1) \\
		\begin{tabular}{crrrr}\\
			& $X_1$ &        \textbf{-1} &         \textbf{0} &         \textbf{1} \\
			\textbf{event} & $X_2$ &           &           &           \\
		    \hline\\
			\multirow{3}{*}{\textbf{\emph{0}}} 
			& \textbf{-1} &   0.364823 &  0.635177 &  0.000000 \\
			&  \textbf{0} &  0.115024 &  0.884976 &  0.000000 \\
			&  \textbf{1} &  0.000000 &  1.000000 &  0.000000 \\
			\cline{1-5}\\
			\multirow{3}{*}{\textbf{\emph{1}}} 
			& \textbf{-1} &  0.476657 &  0.523343 &  0.000000 \\
			&  \textbf{0} &  0.410729 &  0.589271 &  0.000000 \\
			&  \textbf{1} &  0.346157 &  0.653843 &  0.000000 \\
			\cline{1-5}\\
			\multirow{3}{*}{\textbf{\emph{2}}} 
			& \textbf{-1} &  0.000000 &  0.702554 &  0.297446 \\
			&  \textbf{0} &  0.000000 &  0.617759 &  0.382241 \\
			&  \textbf{1} &  0.000000 &  0.533823 &  0.466177 \\
		    \hline
		 \end{tabular}
	\end{tabular}
	\end{scriptsize}		 
	\caption{Price impact with low rate of walking the book and no clustering}
	\label{tab.low_walking}
\end{table}

\begin{figure}[H]
\begin{center}
\includegraphics[width=0.6\textwidth]{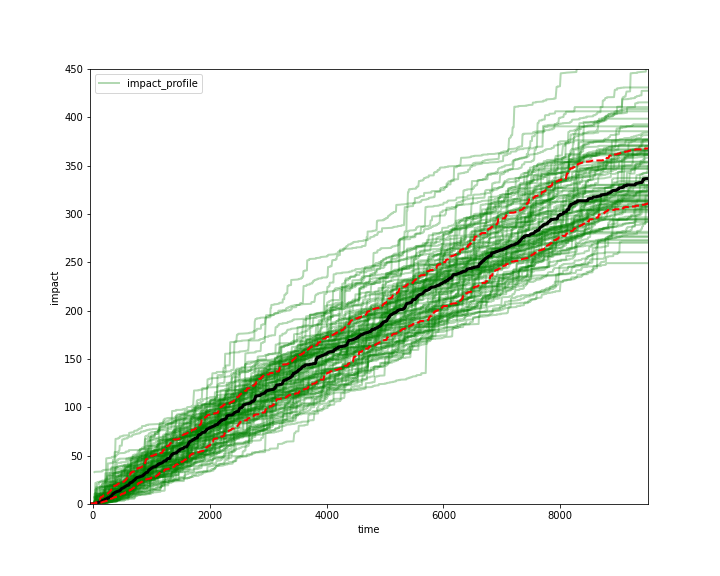}
\caption{Simulated trajectories of the impact profile for the scenario considered in Table \ref{tab.low_walking}. The black solid line is the median trajectory of the impact profile across simulations and the red dotted lines are the 25\% and 75\% quantile trajectories.  }
\label{fig:low_walking_100_sims}
\end{center}
\end{figure}

We remark that the stylised features one observes in the impact profile trajectory are consistent across simulations. In Figure \ref{fig:low_walking_100_sims} we show the impact profile trajectories for one hundred simulations.  

\begin{remark}
Taking the mean across simulations provides a bridge 
from our scenario-dependent impact profiles
to the average impact profiles studied in the literature.
Since our model is rich enough to comprise all 
market variables appearing 
in the stylized facts of price impact,
one can hold all model parameters fixed except for one variable of interest, 
and study average price impact scores as a function of the variable of interest. 
This opens the door to investigating
whether a specific model adheres to the stylized facts in the literature, 
as those reviewed in 
\cite{ZTFL15bey}.
\end{remark}

Table \ref{tab.high_walking}
shows a liquidation in which
the execution schedule is as in Table \ref{tab.low_walking}, 
namely no clustering and same base rate. 
However, in this case liquidator's orders have a much bigger size. 
They have size equal to 50\% of the volumes
of bid offers queuing on the first $n$ levels of the bid side
at the moment of the order submission. 
This entails that liquidator's orders will 
likely walk the book.
Indeed, they 
walk the book whenever the aggregate size of level 1 is 
less than 50\% of the overall bid volume. 
As a result, the estimated transition matrix 
$\transProb_{0} = \transProb_{0}(x\derivative,x)$
concentrates the mass on those states $x=(x_1, x_2)$ such that 
$x_2 = -1$.
The more negative the queue imbalance 
at the moment of the order submission is,
the more this concentration holds. 

The liquidation takes approximately 1100 seconds to complete. 
This is considerably shorter than in the previous simulation 
because every order executes more of the liquidator's inventory.
The price impact score, 
is 6.24\%. 
The plotted mid-price trajectory shows a sharper plunge during execution.

\begin{table}[H]
	\begin{scriptsize}
	\begin{tabular}{c}
		\begin{tabular}{lr}
			Initial inventory $Q_0$: & 10.0 \\
			Base rate $\baseRate_0$: & 0.03 \\
			Clustering rate $\clusteringRate$: & 0.0 \\
			Order size $\ordersizing$: & 0.5 \\
			Start time $t_0$: & 0.0\\
			Termination time $\tau$: & 1104.6\\
			Price impact score: & 0.06244\\
		\end{tabular}
	\\
	\\
		\includegraphics[width=0.45\textwidth]{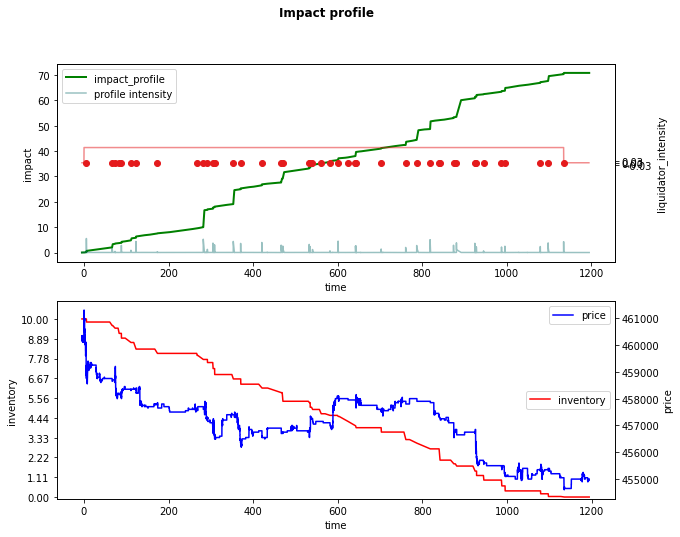}
		\includegraphics[width=0.45\textwidth]{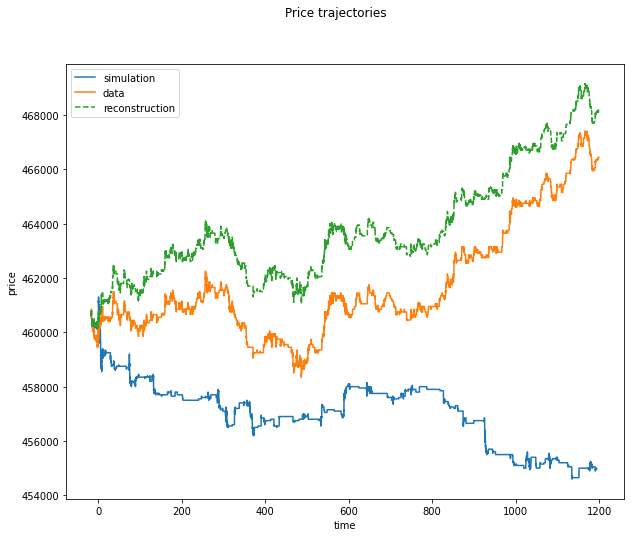}
	\\
	\\	
	\\	
	\\	
        \\
		Probabilities of mid-price movements $X_1(T)\in \lbrace-1, 0, +1\rbrace$ 
		when event of type  0, 1 or 2 occurs at  \\
		time $T$ and 
		queue imbalance $X_2(T-)$ is negative (-1), neutral (0) or positive (+1) \\
		\begin{tabular}{crrrr}\\
			& $X_1$ &        \textbf{-1} &         \textbf{0} &         \textbf{1} \\
			\textbf{event} & $X_2$ &           &           &           \\
		    \hline\\
			\multirow{3}{*}{\textbf{\emph{0}}} 
			& \textbf{-1} &   1.000000 &  0.000000 &  0.000000 \\
			&  \textbf{0} &  0.702381 &  0.297619 &  0.000000 \\
			&  \textbf{1} &  0.666667 &  0.333333 &  0.000000 \\
			\cline{1-5}\\
			\multirow{3}{*}{\textbf{\emph{1}}} 
			& \textbf{-1} &  0.476657 &  0.523343 &  0.000000 \\
			&  \textbf{0} &  0.410729 &  0.589271 &  0.000000 \\
			&  \textbf{1} &  0.346157 &  0.653843 &  0.000000 \\
			\cline{1-5}\\
			\multirow{3}{*}{\textbf{\emph{2}}} 
			& \textbf{-1} &  0.000000 &  0.702554 &  0.297446 \\
			&  \textbf{0} &  0.000000 &  0.617759 &  0.382241 \\
			&  \textbf{1} &  0.000000 &  0.533823 &  0.466177 \\
		    \hline
		 \end{tabular}
	\end{tabular}
	\end{scriptsize}		 
	\caption{Price impact with high rate of walking the book and no clustering}
	\label{tab.high_walking}
\end{table}

Hence, Tables \ref{tab.low_walking} and \ref{tab.high_walking}
show how our model captures
the consequence that order sizing has on price impact.
Order sizing however is not the main driver of price impact.
We demonstrate this in Table \ref{tab.with_the_market}.

Table \ref{tab.with_the_market}
shows a liquidation in which 
the intensity of the execution schedule
has zero base rate, 
namely there is no exogenous cause 
for the liquidator to submit their orders.
Instead, 
they submit their orders in response to
orders submitted by other market participants. 
We set the liquidator's response to be proportional
to the response that other market participants 
make when scheduling their sell market orders. 
That is, we set 
$\impCoef_{\eone, x, 0} = \clusteringRate \cdot \impCoef_{\eone, x, 1}$,
for some coefficient $\clusteringRate \geq 0$. 
In the simulation of Table \ref{tab.with_the_market} this coefficient is 
$\clusteringRate = 0.25$.

Recall that $\impCoef_{\eone, x, 1}$ was estimated by 
maximum likelihood estimation
from the LOBSTER data set -- 
see Table \ref{tab.estimatedHawkesParam} 
where the $\Lone$-norms 
$\impCoef_{\eone, x, e} / (\decCoef_{\eone, x, e} - 1)$ 
are reported.
The estimation revealed that 
seller-initiated executions 
are more likely to excite other events with deflationary pressure on the mid-price,
i.e. events of type $e=1$ or $e=3$,
whereas
buyer-initiated executions 
are more likely to excite other events with inflationary pressure on the mid-price,
i.e. events of type $e=2$ or $e=4$.
As a consequence, 
when $\impCoef_{\eone,x,0}$ is proportional to 
$\impCoef_{\eone, x, 1}$
deflationary events will tend to cluster,
potentially triggering abrupt price changes.
The line charts in Table \ref{tab.with_the_market}
give visual representations of this. 
We deliberately focused on a short time horizon of 300 seconds,
so that the phenomenon described is more apparent to the eye.

We set the order sizes to be very small, at 1.5\% of the 
volumes available on the first $n$ levels of the bid side. 
This has the purpose to isolate the phenomenon of
indirectly induced price changes versus 
price changes induced by walking the book. 
Indeed, the estimated transition probabilities $\transProb_{0}$ on this simulation
assign negligible probabilities to transitioning
to a state $\stateVariable(T)$ with $\stateVariable_1(T) = -1$ when 
an event of type 0 occurs at time $T$. 

Nonetheless, because of the clustering effect of deflationary events,
the price impact score is the highest, at 14.08\%.

\begin{table}[H]
	\begin{scriptsize}
	\begin{tabular}{c}
		\begin{tabular}{lr}
			Initial inventory $Q_0$: & 10.0 \\
			Base rate $\baseRate_0$: & 0.0 \\
			Clustering rate $\clusteringRate$: & 0.25 \\
			Order size $\ordersizing$: & 0.015 \\
			Start time $t_0$: & 0.0\\
			Termination time $\tau$: & 5201.3\\
			Price impact score: & 0.1408\\
		\end{tabular}
	\\
	\\
		\includegraphics[width=0.45\textwidth]{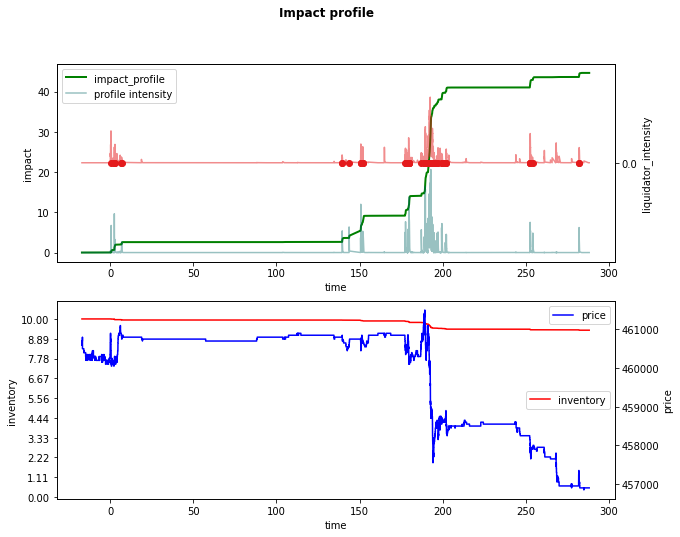}
		\includegraphics[width=0.45\textwidth]{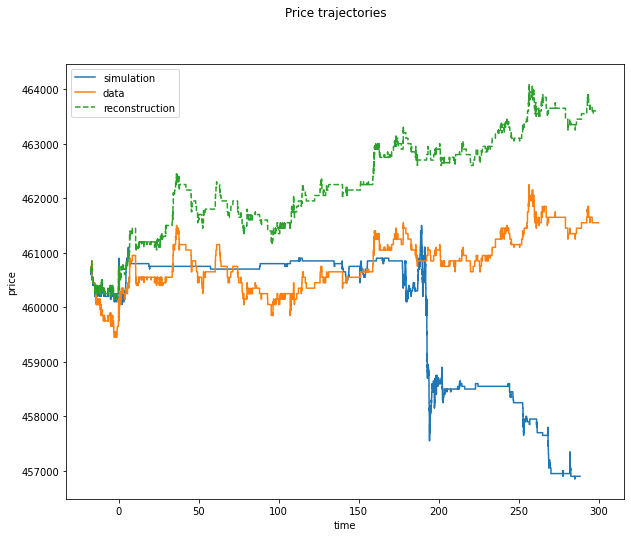}
	\\
	\\	
	\\	
	\\	

		Probabilities of mid-price movements $X_1(T)\in \lbrace-1, 0, +1\rbrace$ 
		when event of type  0, 1 or 2 occurs at  \\
		time $T$ and 
		queue imbalance $X_2(T-)$ is negative (-1), neutral (0) or positive (+1) \\
		\begin{tabular}{crrrr}\\
			& $X_1$ &        \textbf{-1} &         \textbf{0} &         \textbf{1} \\
			\textbf{event} & $X_2$ &           &           &           \\
		    \hline\\
			\multirow{3}{*}{\textbf{\emph{0}}} 
			& \textbf{-1} &   0.074063 &  0.925937 &  0.000000\\
			&  \textbf{0} &  0.007691 &  0.992309 &  0.000000\\
			&  \textbf{1} &  0.000000 &  1.000000 &  0.000000  \\
			\cline{1-5}\\
			\multirow{3}{*}{\textbf{\emph{1}}} 
			& \textbf{-1} &  0.476657 &  0.523343 &  0.000000 \\
			&  \textbf{0} &  0.410729 &  0.589271 &  0.000000 \\
			&  \textbf{1} &  0.346157 &  0.653843 &  0.000000 \\
			\cline{1-5}\\
			\multirow{3}{*}{\textbf{\emph{2}}} 
			& \textbf{-1} &  0.000000 &  0.702554 &  0.297446 \\
			&  \textbf{0} &  0.000000 &  0.617759 &  0.382241 \\
			&  \textbf{1} &  0.000000 &  0.533823 &  0.466177 \\
		    \hline
		 \end{tabular}
	\end{tabular}
	\end{scriptsize}		 
	\caption{Price impact with clustering of liquidator's orders and low rate of walking the book}
	\label{tab.with_the_market}
\end{table}

The results presented in this section are robust to misspecification of model parameters. When we stress the base rates $\{\nu_e\}_{e}$, together with the coefficients $\{\alpha_{e',x',e}\}_{e',x',e}$ and $\{\beta_{e',x',e}\}_{e',x',e}$ with shocks between $-5\%$ and $+5\%$, we observe that the average price impact scores across one hundred simulations get affected by less than $10\%$ of the average unstressed values. For example, if we consider the experiment in Table \ref{tab.low_walking}, and we repeat the profiling with the modified parameters $\tilde{\nu}_e = 1.05\,\nu_e$, $\tilde{\alpha}_{e',x',e} = 1.05\,\alpha_{e',x',e}$ and $\tilde{\beta}_{e',x',e} = 1.05\,\beta_{e',x',e}$ for all $\{e',x,e\}$, we find that the average price impact scores changes from 0.037466 (with 0.006 standard deviation) to 0.04055 (with 0.006 standard deviation), which represents an average increase of $8.2\%$. Similarly, in that same scenario of Table \ref{tab.low_walking}, when we consider the modified parameters $\tilde{\nu}_e = 0.95\,\nu_e$, $\tilde{\alpha}_{e',x',e} = 0.95\,\alpha_{e',x',e}$ and $\tilde{\beta}_{e',x',e} = 0.95\,\beta_{e',x',e}$ for all $\{e',x,e\}$, we find that the average price impact scores decrease to 0.036062 (with 0.005) standard deviation), which represents an average decrease of $3.7\%$. The other two scenarios we consider in Tables \ref{tab.high_walking} and \ref{tab.with_the_market} show a similar behaviour when stressing the calibrated parameters. This means that the results shown are indeed robust to calibration errors.

Finally, we repeat the above analysis employing the tickers AAPL and INTC for a range of dates during January 2019. We report that no further insights are derived from considering  other dates or other tickers. For example, it was always the case that the highest price impact scores were achieved under the scenario we consider in Table \ref{tab.with_the_market}, i.e., where the clustering effect plays a key role in the price impact profile. Thus, we conclude that the intuition derived from the analysis in Tables \ref{tab.low_walking}, \ref{tab.high_walking}, and \ref{tab.with_the_market} remains valid for other dates and similar tickers. 

In limit order books of small tick-size stocks (e.g, GOOG or TSLA), 
orders are posted and cancelled in adjacent price queues more frequently than 
for medium-size or large-size stocks.
In other words, the mid-price of a small tick-size stock has far smaller constant traits. 
This is because, relative to the stock price, a one-tick change in the best bid or best ask is not as significant as for larger tick-size stocks. 
When studying small tick-size stocks with our model,
such a reduced importance of single-tick changes can be accounted for
by letting $\countingProc[3]$ and $\countingProc[4]$ 
jump only when the mid-price changes by two or more ticks,
de-facto renormalising the parameters of the limit order book (doubling or tripling the tick size and merging adjacent queues of orders). 
In our experiments, this led to a more robust calibration  and prevented overfitting.


\clearpage
\newpage
\addcontentsline{toc}{section}{Bibliography}
\bibliographystyle{apalike}
\bibliography{bibliography}
\begin{appendices}
\section{Proofs}\label{sec.proofs}

\begin{proof}[Proof of Propostion \ref{prop.decomposition_of_limit_order}]
Let $(t,q,p,-1)$ be a sell limit order.
Let
$N_v := \inf\lbrace n\geq 0 : \, q<\sum_{i=1}^{n} \nthBestBidVolume[i]_{t-}\rbrace$.
Let 	
$N_p:=\inf\lbrace n\geq 1: \,  \nthBestBidPrice[n]_{t-}< p \rbrace$.
Let
\begin{equation*}
\begin{split}
q^{i}:=& \max\left( 0, 
q-\sum_{k=1}^{i\wedge N_v \wedge (N_p - 1)} \nthBestBidVolume[k]_{t-}
\right)
\\
=& \max\left( 0, 
q-\sum_{1 \leq k\leq i} \nthBestBidVolume[k]_{t-}\one\left\lbrace \nthBestBidPrice[k]_{t-} \geq p\right\rbrace
\right)
\end{split}
\end{equation*}
Let $N_v^M$, $N_p^M$ and $q_M^i$ be the corresponding quantities for the order  $(t,q_M,0,-1)$.
Notice that $N_v^M = N_v^M \wedge N_p^M$.
Processing $(t,q_M,0,-1)$ 
does not affect the ask side 
because $q^{\infty}_M = 0$; 
moreover the effects on the bid side of 
processing $(t,q_M,0,-1)$ 
are the same as those of $(t,q,p,-1)$, 
because 
$\inf \lbrace n\geq 0 : \, q_M < \sum_{i=1}^{n} \nthBestBidVolume[i]_{t-} \rbrace = N_v \wedge N_p$ and $q^{k+N-1} - q^{k+N} = q_M^{k+N_v^M - 1} - q_M^{k+N_v^M}$.
Therefore, 
after $(t,q_M,0,-1)$ has been processed, 
the bid side is the same as the bid side 
after the processing of $(t,q,p,-1)$.

Processing $(t,q-q_M,p,-1)$ 
after $(t,q_M,0,-1)$ 
does not alter the bid side 
because either $\bestBidPrice_{t-1} < p$ or $q-q_M=0$.
Moreover, 
$(t,q-q_M,p,-1)$ has the same effects on the ask side 
as those of $(t,q,p,-1)$ because $q-q_M = q^{\infty}$.

The proof of the decomposition of a buy limit order is mutatis mutandis the same.
\end{proof}

\begin{proof}[Proof of Proposition \ref{prop.sufficient_cond_price_symmetry}.]
 Let $(\arrivalTimes[]_n,\event[n], \stateVariable_n)$ be the sequence of arrival times, events and states.
Then, we can compute
 \begin{equation*}
  \begin{split}
   \sum_{x\in\inflationarySpace}\sum_{e=1}^{4}
   \transProb_{e}(\stateVariable(t),& x)
   \sum_{\eone=1}^{4}\int_{[0,t)}\hawkesKernel\subscriptee(t-s,\stateVariable(s))d\countingProc[\eone](s)
   \\
   =&
   \sum_{n:\arrivalTimes[]_n <t}
   \sum_{x\in\inflationarySpace}\sum_{e=1}^{4}
   \transProb_{e}(\stateVariable(t),x)
   \hawkesKernel_{\event[n],e}(t-\arrivalTimes[]_n,\stateVariable_n)
   \\
   =&
   \sum_{n:\arrivalTimes[]_n <t}
   \sum_{x\in\inflationarySpace}\sum_{e=1}^{4}
   \transProb_{\sigma_{E}(e)}(\stateVariable(t),\sigma_{\stateSpace}(x))
   \hawkesKernel_{\event[n],\sigma_{E}(e)}(t-\arrivalTimes[]_n,\stateVariable_n)
   \\
   =&
   \sum_{n:\arrivalTimes[]_n <t}
   \sum_{x\in\deflationarySpace}\sum_{e=1}^{4}
   \transProb_{e}(\stateVariable(t),x)
   \hawkesKernel_{\event[n],e}(t-\arrivalTimes[]_n,\stateVariable_n)
   \\   
   =&
   \sum_{x\in\deflationarySpace}\sum_{e=1}^{4}
   \transProb_{e}(\stateVariable(t), x)
   \sum_{\eone=1}^{4}\int_{[0,t)}\hawkesKernel\subscriptee(t-s,\stateVariable(s))d\countingProc[\eone](s).
   \end{split}
 \end{equation*}
\end{proof}

\begin{proof}[Proof of Proposition \ref{prop.bm_impact}.]
	We need to show that
 \begin{equation}\label{eq.intensity_of_BM-impact}
 \Big(\intensityDeflationary - \intensityInflationary\Big)(t)
  =\directImpact(t) + \indirectImpact(t).
  \end{equation}
  To this purpose, we compute 
$\intensityDeflationary$ 
(resp., of $\intensityInflationary$) 
as the sum of the intensities of $\hybridHawkes_{e,x}$ 
for $e = 0,\dots,4$ and $x$ in $\deflationarySpace$ 
(resp., in $\inflationarySpace$). 
From equations \eqref{eq.intensity_of_hybridMPP} 
and \eqref{eq.intensities_when_liquidator_is_present} 
it follows that 
 \begin{equation*}
 \begin{split}
  \intensityDeflationary(t) 
  =&
  \sum_{x\in\deflationarySpace}\Bigg\lbrace
  \transProb_{0}(\stateVariable(t),x) \intensity[0](t)
  \\
  &  +
  \sum_{e=1}^{4} \transProb_{e}(\stateVariable(t),x) 
  \Bigg(
  \underbrace{
  \baseRate_e 
 + \sum_{\eone = 1}^{4}\sum_{\xone=1}^{3K}\int_{[0,t)} \hawkesKernel_{\eone,e}(t-s,\xone)d\hybridHawkes_{\eone,\xone}(s)
 }_{=\ell_{e}(t)}
 + \int_{[0,t)} \hawkesKernel_{1,e}(t-s,\xone)d\hybridHawkes_{0,\xone}(s)
 \Bigg)
  \Bigg\rbrace,
  \\
  \end{split}
 \end{equation*}
 where $\intensity[0](t)$ is as in \eqref{eq.intensity_of_liquidator}, and 
 \begin{equation*}
 \begin{split}
  \intensityInflationary(t) 
  =&
  \sum_{x\in\inflationarySpace}
  \sum_{e=1}^{4} \transProb_{e}(\stateVariable(t),x) 
  \Bigg(
  \underbrace{
  \baseRate_e 
 + \sum_{\eone = 1}^{4}\sum_{\xone=1}^{3K}\int_{[0,t)} \hawkesKernel_{\eone,e}(t-s,\xone)d\hybridHawkes_{\eone,\xone}(s)
 }_{=\ell_{e}(t)}
 + \int_{[0,t)} \hawkesKernel_{1,e}(t-s,\xone)d\hybridHawkes_{0,\xone}(s)
 \Bigg).
   \end{split}
 \end{equation*}
 By price-symmetry, the terms $\sum_{x\in\deflationarySpace}\sum_{e=1}^{4}\transProb_{e}(\stateVariable(t),x) \ell_{e}(t)$ and $\sum_{x\in\inflationarySpace}\sum_{e=1}^{4}\transProb_{e}(\stateVariable(t),x) \ell_{e}(t)$ will  cancel out from the difference $\intensityDeflationary(t) - \intensityInflationary(t)$. 
 \end{proof}

\end{appendices}
\end{document}